\documentclass[utf8]{frontiersSCNS} 

\usepackage{url,hyperref,lineno,microtype,subcaption}
\usepackage[onehalfspacing]{setspace}

\def\keyFont{\fontsize{8}{11}\helveticabold }
\def\firstAuthorLast{Gobrecht} 
\def\Authors{David Gobrecht\,$^{1,*}$}
\def\Address{$^{1}$Institute of Astronomy, Department of Physics and Astronomy, KU Leuven, Leuven, Belgium} 
\def\corrAuthor{David Gobrecht}

\def\corrEmail{dave@gobrecht.ch}

\begin{document}
\onecolumn
\firstpage{1}

\title[(SiC)\textsubscript{n} ionization]{The vertical and adiabatic ionization energies of silicon carbide clusters, (SiC)\textsubscript{n}, with n=1-12} 

\author[\firstAuthorLast ]{\Authors} 
\address{} 
\correspondance{} 

\extraAuth{}

\maketitle

\begin{abstract}
Silicon carbide (SiC) is one of the major cosmic dust components in carbon-rich environments.
However, the formation of SiC dust is not well understood. In particular, the initial stages of the SiC condensation (i.e. the SiC nucleation) remain unclear, as the basic building blocks (i.e. molecular clusters) exhibit atomic segregation at the (sub-)nanoscale. 
We report vertical and adiabatic ionization energies of small silicon carbide clusters, (SiC)\textsubscript{n}, n=2$-$12, ranging from 6.6$-$10.0 eV, which are lower than for the SiC molecule ($\sim$10.6 eV). 
The most favorable structures of the singly ionized (SiC)\textsubscript{n}\textsuperscript{+}, n=5$-$12, cations resemble their neutral counterparts. However, for sizes n=2$-$4, these structural analogues are metastable and different cation geometries are favored. Moreover, we find that the (SiC)\textsubscript{5}\textsuperscript{+} cation is likely to be a transition state.
Therefore, we place constraints on the stability limit for small, neutral (SiC)\textsubscript{n} clusters to persist ionization through (inter)-stellar radiation fields or high temperatures.
\section{}


\tiny
 \keyFont{ \section{Keywords:} silicon carbide, SiC, ionization, clusters, cations, nucleation, dust formation, circumstellar} 
\end{abstract}

\section{Introduction}
The chemistry occurring in astrophysical environments is primarily controlled by the carbon-over-oxygen (C/O) ratio. This is a consequence of the high carbon monoxide (CO) bond energy of 11.2 eV making CO the most stable diatomic molecule known \citep{luo2007comprehensive}.
This classical dichotomy is challenged since non-equilibrium processes such as photo-chemistry and
 pulsation-induced shocks can break the strong CO bond and lead to the formation of unexpected molecules \citep{Ag_ndez_2010,2016A&A...585A...6G}. Of particular astronomical interest are refractory molecules and molecular clusters that represent the precursors of circumstellar dust.\   
One of the main dust species in carbon-dominated regimes is silicon carbide (SiC). 
A broad spectral feature around 11.3 micron is commonly observed in Carbon-rich evolved starts and attributed to the presence of SiC dust grains \citep{https://doi.org/10.1002/asna.19682910405,1972A&A....21..239H,1974ApJ...188..545T}.
SiC stardust was extracted from pristine meteorites \citep{1987Natur.330..728B,1994GeCoA..58..459A,HOPPE1996883,2007GeCoA..71.4786Z,0004-637X-786-1-66}. Recent studies have shown that the vast majority of presolar SiC grains found in pristine meteoritic stardust originate from low-mass Asymptotic Giant Branch (AGB) stars \citep{2020A&A...644A...8C}. But also molecular gas-phase species like SiC, Si\textsubscript{2}C, SiC\textsubscript{2} were detected in the circumstellar envelopes of carbon-rich evolved stars \citep{1989ApJ...341L..25C,doi:10.1021/acs.jpclett.5b00770,1984ApJ...283L..45T,massalkhi}. The evidence for both, gas-phase silicon-carbon molecules and solid SiC dust suggests that their intermediaries, i.e. SiC molecular clusters, also exist in carbon-rich astronomical environments and partake in the nucleation and SiC dust formation process.\\ 
Therefore, SiC molecular clusters are the objects of our interest.
This study represents a continuation of a prior work \citep{2017ApJ840117G} and adresses the (single) ionization energies of the previously investigated neutral (SiC)\textsubscript{n}, n=1-12, clusters.
This paper is organized as follows.  
In Section 2 we present the methods used to derive the vertical and adiabatic ionization energies. Section 3 shows the results for these energies as well as the adiabatically optimized cation geometries and in Section 4 our summary and conclusion is given.


\section{Methods}
In this study, we focus on small, singly ionized silicon carbide cation clusters with a silicon-to-carbon stoichiometry of 1:1, (SiC)\textsubscript{n}\textsuperscript{+}. 
The smallest lowest-energy neutral (SiC)\textsubscript{n}, n=1$-$11, clusters are characterized by atomic segregation of carbon and silicon \citep{:/content/aip/journal/jcp/128/15/10.1063/1.2895051,molecules18078591,:/content/aip/journal/jcp/145/2/10.1063/1.4955196}. The carbon atoms tend to form chains and benzene-like rings, whereas the silicon atoms
develop triangular and tetrahedral substructures bridging to the carbon segregation. The presence of aromatic carbon rings in the neutral (SiC)\textsubscript{n} Global Minima (GM) candidates indicates the occurence of strongly bound, delocalised valence electrons .\\     
We investigate the vertical and adiabatic single-electron ionization energies of (SiC)\textsubscript{n}, n=1$-$12, clusters by means of Density Functional Theory (DFT). The energies and geometries of the lowest-energy neutral isomers (i.e. GM candidates) were already presented in a previous study \citep{2017ApJ840117G}.
These calculations were performed on the M11/cc-pVTZ level of theory \citep{doi:10.1021/jz201525m}, which was shown to result in accurate energies with affordable computational cost and, in contrast to the B3LYP functional, is not outperforming \citep{:/content/aip/journal/jcp/145/2/10.1063/1.4955196}. 
For consistency and comparability, the (SiC)\textsubscript{n}\textsuperscript{+} cluster calculations were also performed on the same level of theory (M11/cc-pVTZ). For the DFT calculations we use the computational software Gaussian 09 \citep{Gaussian09}.
For all optimizations a vibrational frequency analysis is performed. First, it allows us to discriminate between Transition States (TSs) exhibiting one imaginary frequency and true minima with only real vibration modes. In the case of a TS, marking a saddle point on the potential energy surface, the geometry is distorted 
in direction of the negative curvature of the saddle point and re-optimized to obtain a real minimum.    
Second, partition functions can be constructed to calculate thermodynamic potentials within the Rigid Rotor Harmonic Oscillator (RRHO) appromixation.  
Third, the resulting vibrational spectra can be compared with observational data.\\  
The current work represents an extension of a previous (SiC)\textsubscript{n} cluster study \citep{2017ApJ840117G}, which did not comprise the ionization of these clusters and its related cations.
Owing to the enormous computational task, we did not perform global optimization searches of the (SiC)\textsubscript{n}\textsuperscript{+} cations themselves.
Instead, we perform ionization calculations for the four lowest-lying neutral isomers for each size n.
As mentioned in \citep{2017ApJ840117G}, the GM candidate isomer may not be the most favorable structure in circumstellar conditions and hence it is required to study a range of the energetically lowest-lying structures and their respective cations for each cluster size. 
Therefore, we assume that the neutral (SiC)\textsubscript{n} clusters are initially present in the form of their four lowest-energy isomers including their GM candidate structure, and that they are subsequently ionized.
We are aware that our set of isomers is not complete and hence, there is the possibility that (SiC)\textsubscript{n} cation isomers with a lower energy exist. 
In order to reduce this possibility and to avoid missing a particularly stable SiC cluster cation, additional test calculations for different geometries (e.g. cages and tubes) are performed.
 

It is known that neutral molecules and related clusters can alter their ground-state geometry significantly upon the loss of an electron. They include, for example, clusters of water \citep{belau2007} and Argon \citep{PhysRevLett.109.193401}, but also organic compounds like methyl ketene \citep{D0CP03921G}, acetic acid \citep{doi:10.1063/1.4754273} or $\delta$-Valerolactam \citep{https://doi.org/10.1002/cphc.201100090}.
Moreover, the spin-degenerate cluster cations can be subject to Jahn-Teller effects implying a lowering of their spatial symmetry and a structural rearrangement. Therefore, it is instructive to not only study the \textit{vertical} ionization energies, corresponding to single point energy evaluations, but also \textit{adiabatic} ionization energies taking into account structural rearrangements.
Here, all cations are presumed to be in a doublet state (i.e. with a spin multiplicity of 2), since the neutral (SiC)\textsubscript{n}, n$\ge$2, clusters are singlet states and the SiC molecule (n=1) is in a triplet state.



\section{Results}
In the following, we present the vertical and adiabatic ionization energies of (SiC)\textsubscript{n}, n=1$-$12, clusters and the related structures of the adiabtically optimized, singly ionized (SiC)\textsubscript{n}\textsuperscript{+}, n=1$-$12, cations.

\subsection{Vertical ionization}
The vertical ionization energies, E\textsubscript{n}\textsuperscript{vert}, are listed in Table \ref{table1}. 
E\textsubscript{n}\textsuperscript{vert} of the lowest-energy cations (denoted with GM) show an overall monotonically decreasing trend with cluster size n, starting from 10.64 eV (n=1) to 6.69 eV (n=12), with opposing tendencies at n=5 and n=9.
For the smallest sizes (n=1$-$2) E\textsubscript{n}\textsuperscript{vert} decrease more strongly than for larger sizes. For the latter size regime the decrease in E\textsubscript{n}\textsuperscript{vert} flattens, except for n=9, where E\textsubscript{9}\textsuperscript{vert}=7.34 is slightly larger than for n=8 and n=10. Considering only structural analogues of the neutral GM clusters (denoted as LM), we find a similar decreasing trend with cluster size n, except for n=9.

\subsection{Adiabatic ionization}
\begin{figure}[h!]
\begin{center}
\includegraphics[width=2cm]{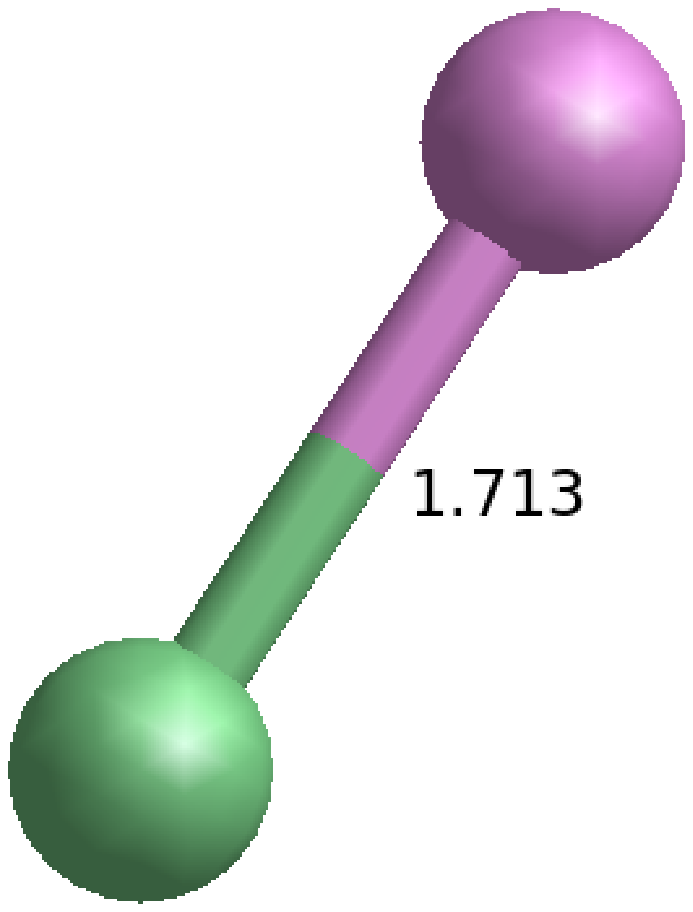}
1
\includegraphics[width=3cm]{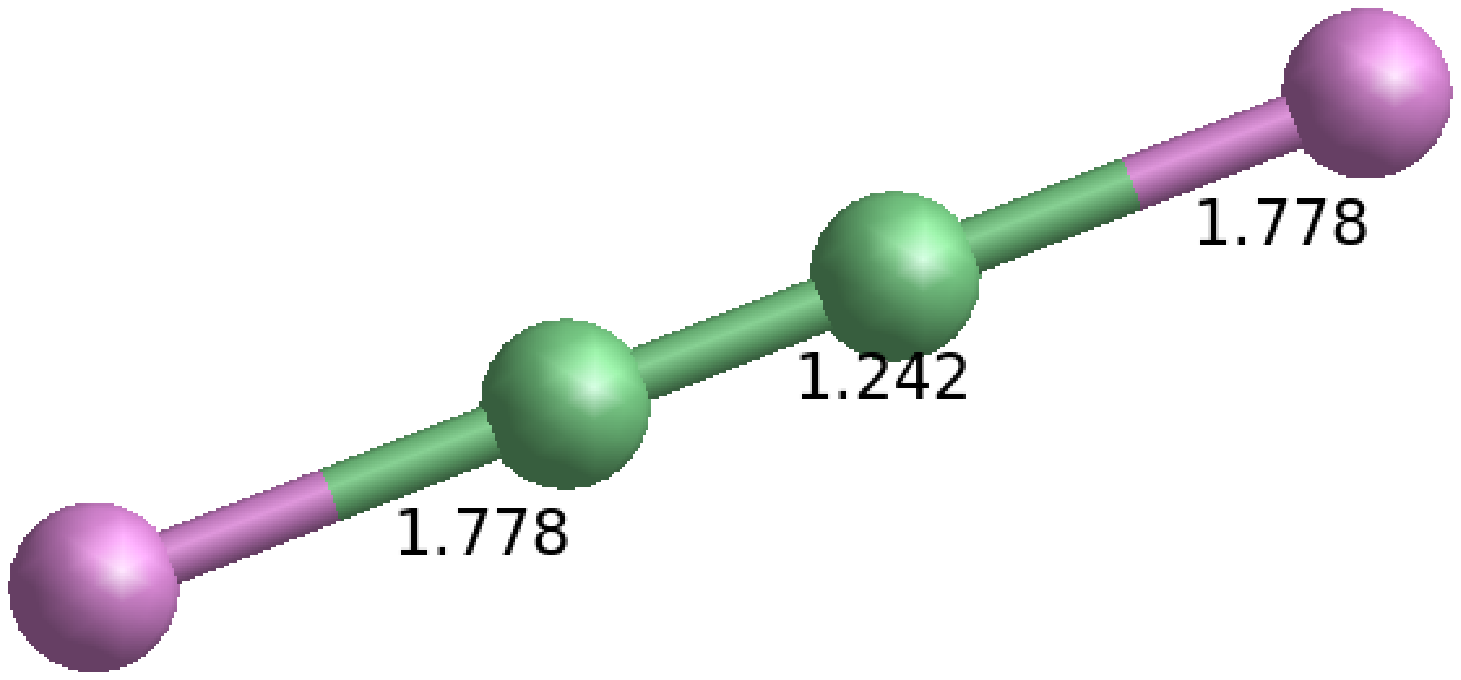}
2
\includegraphics[width=3cm]{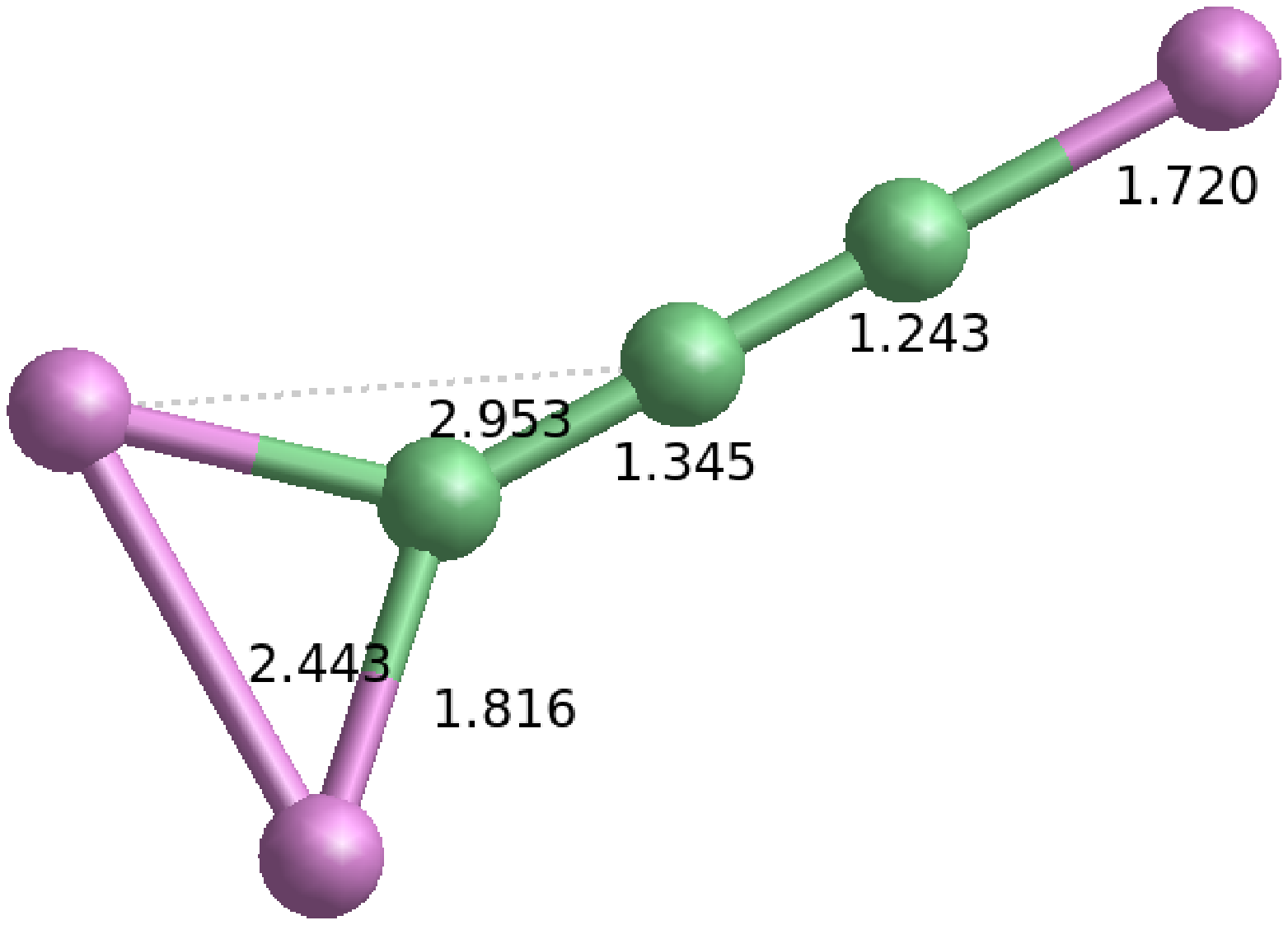}
3
\includegraphics[width=3cm]{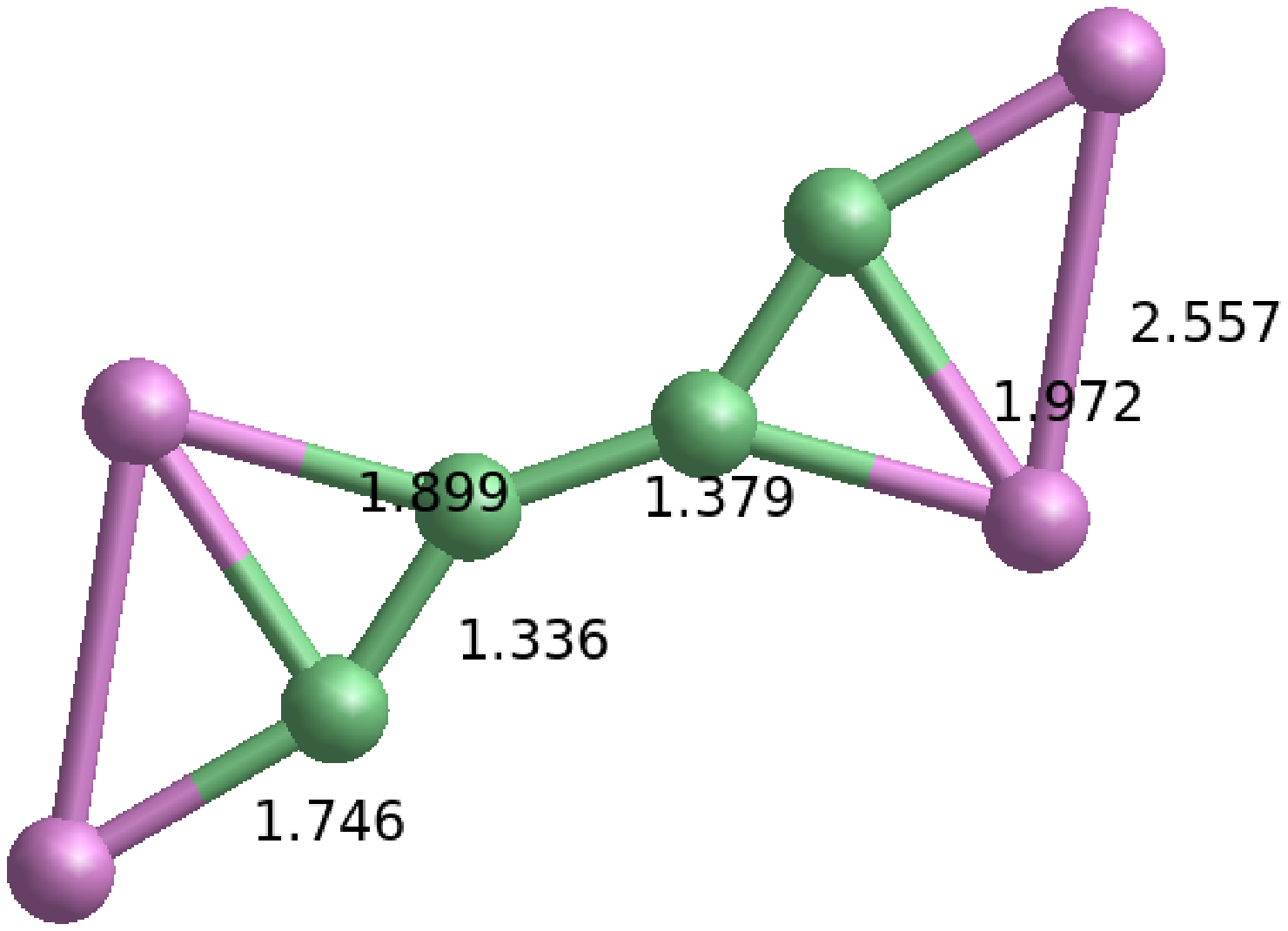}
4
\includegraphics[width=3cm]{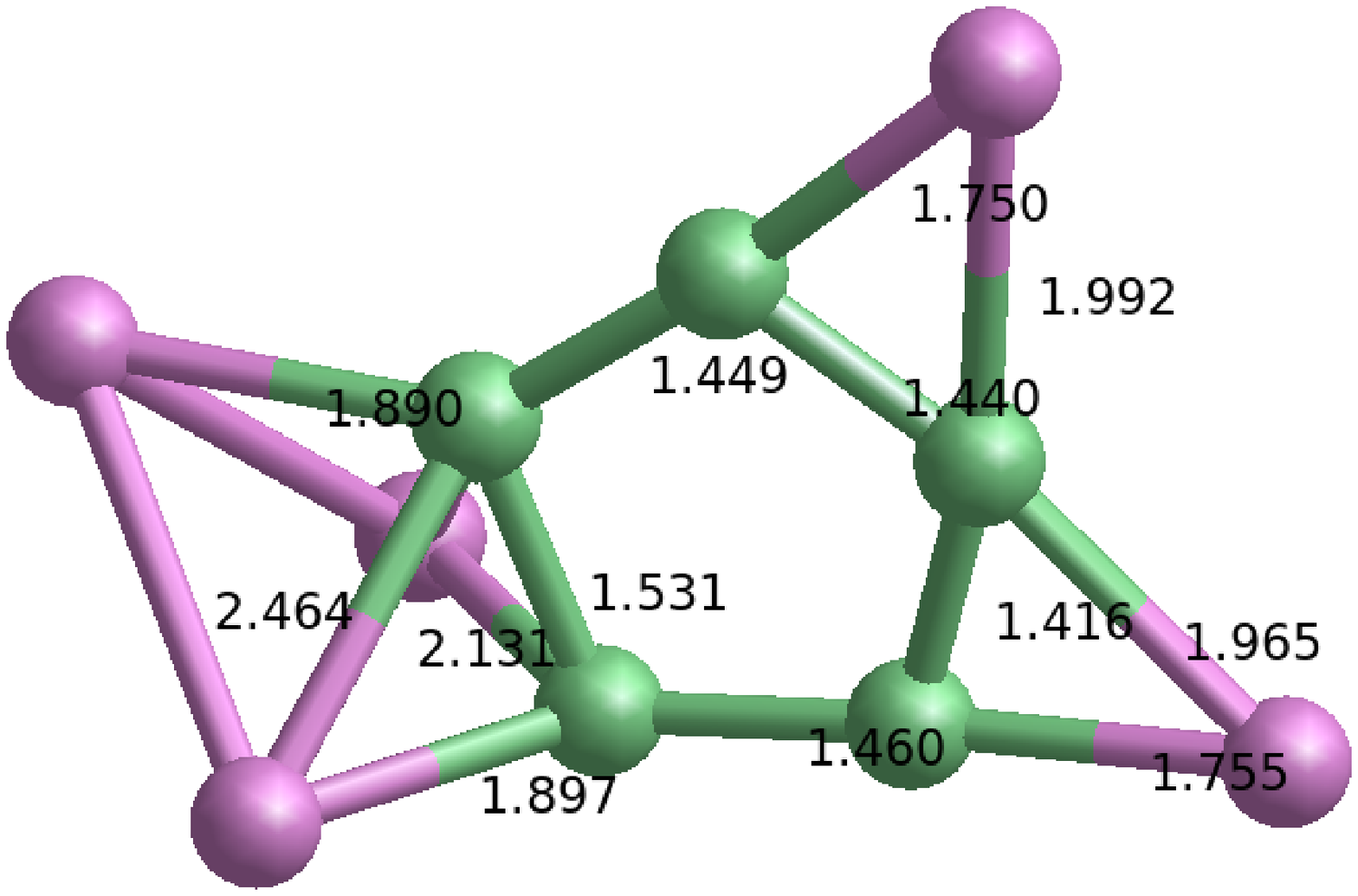}
5(TS)
\includegraphics[width=3cm]{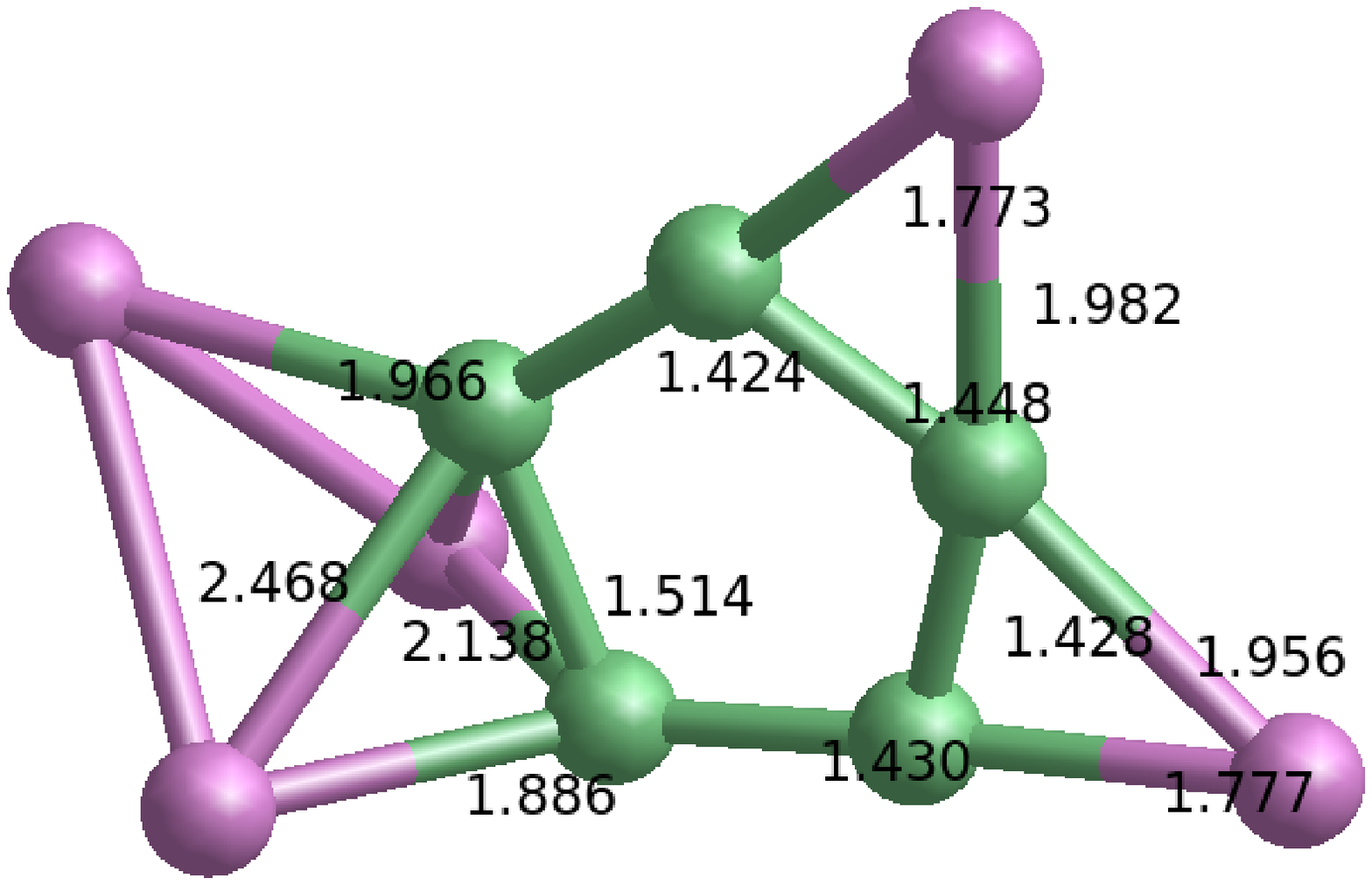}
5(GM)
\includegraphics[width=3cm]{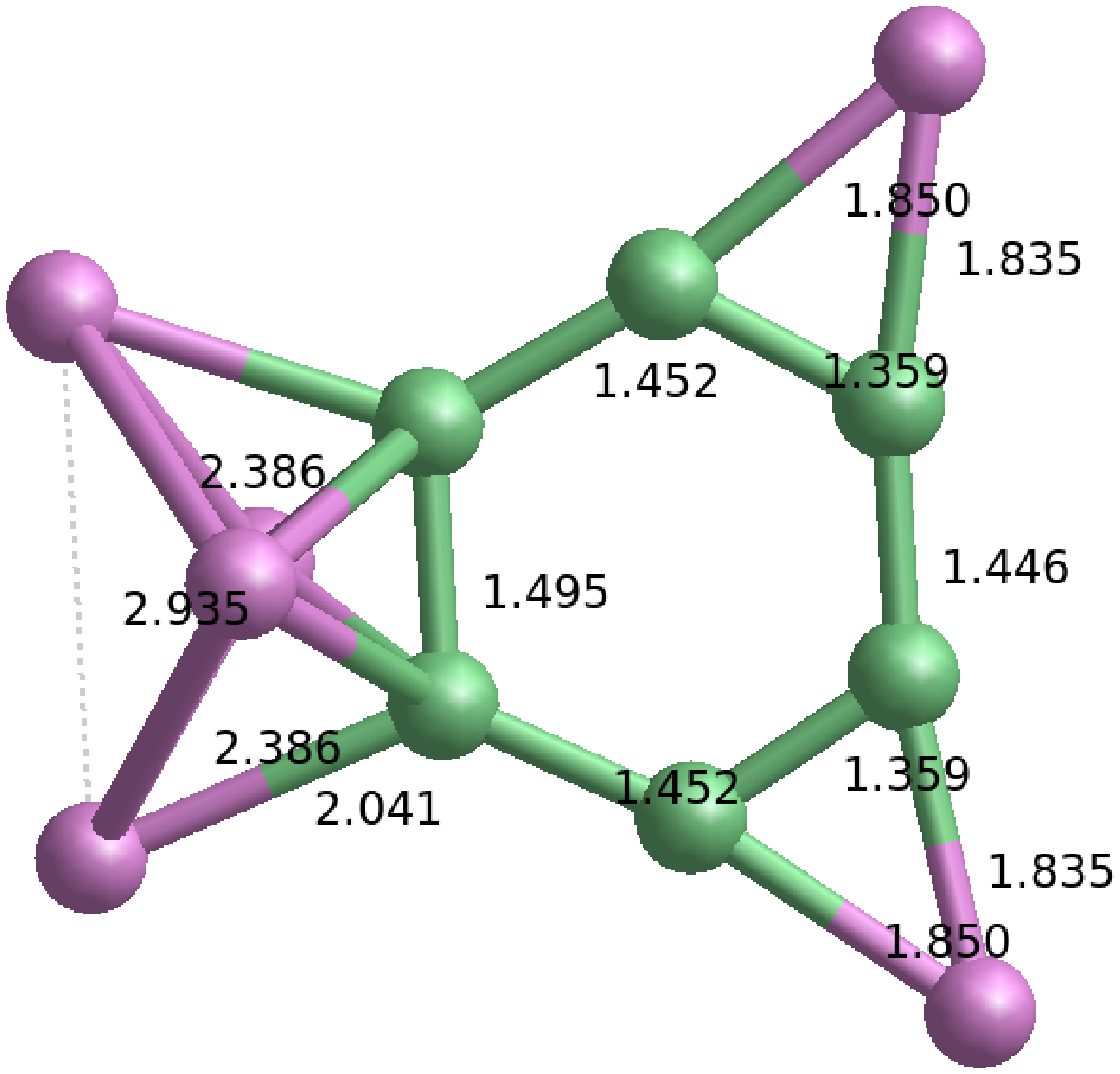}
6
\includegraphics[width=3.5cm]{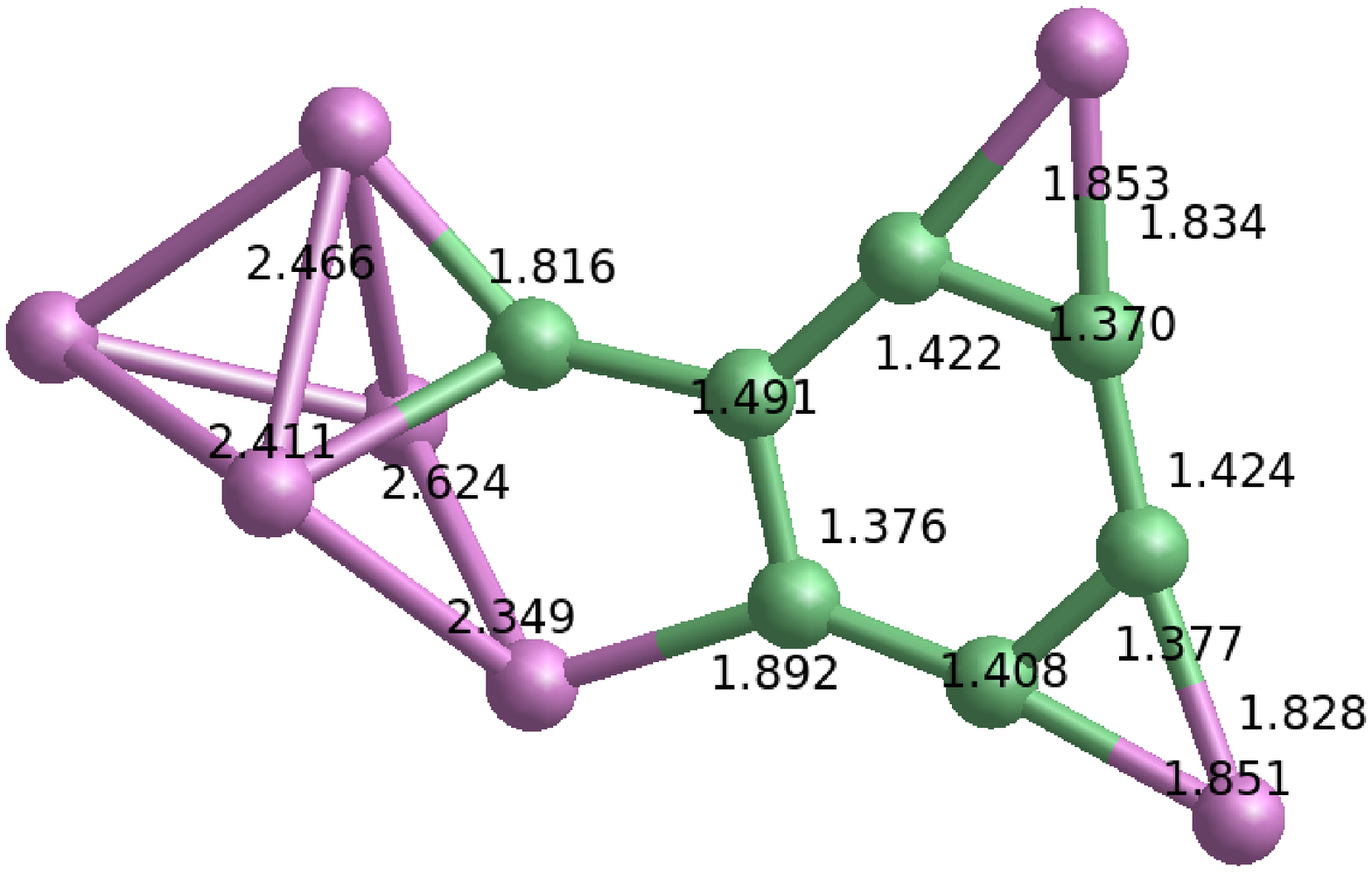}
7
\includegraphics[width=3.5cm]{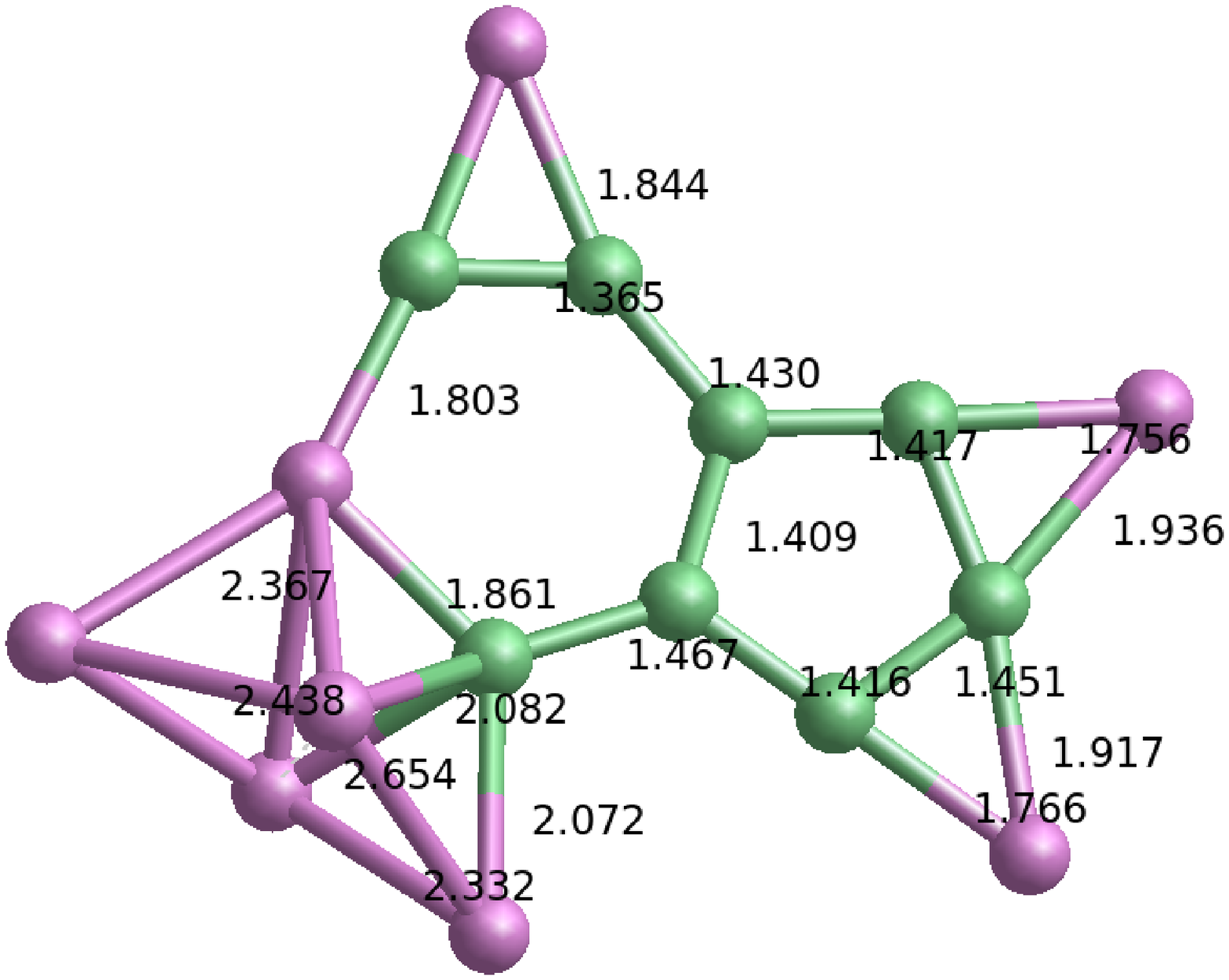}
8
\includegraphics[width=4cm]{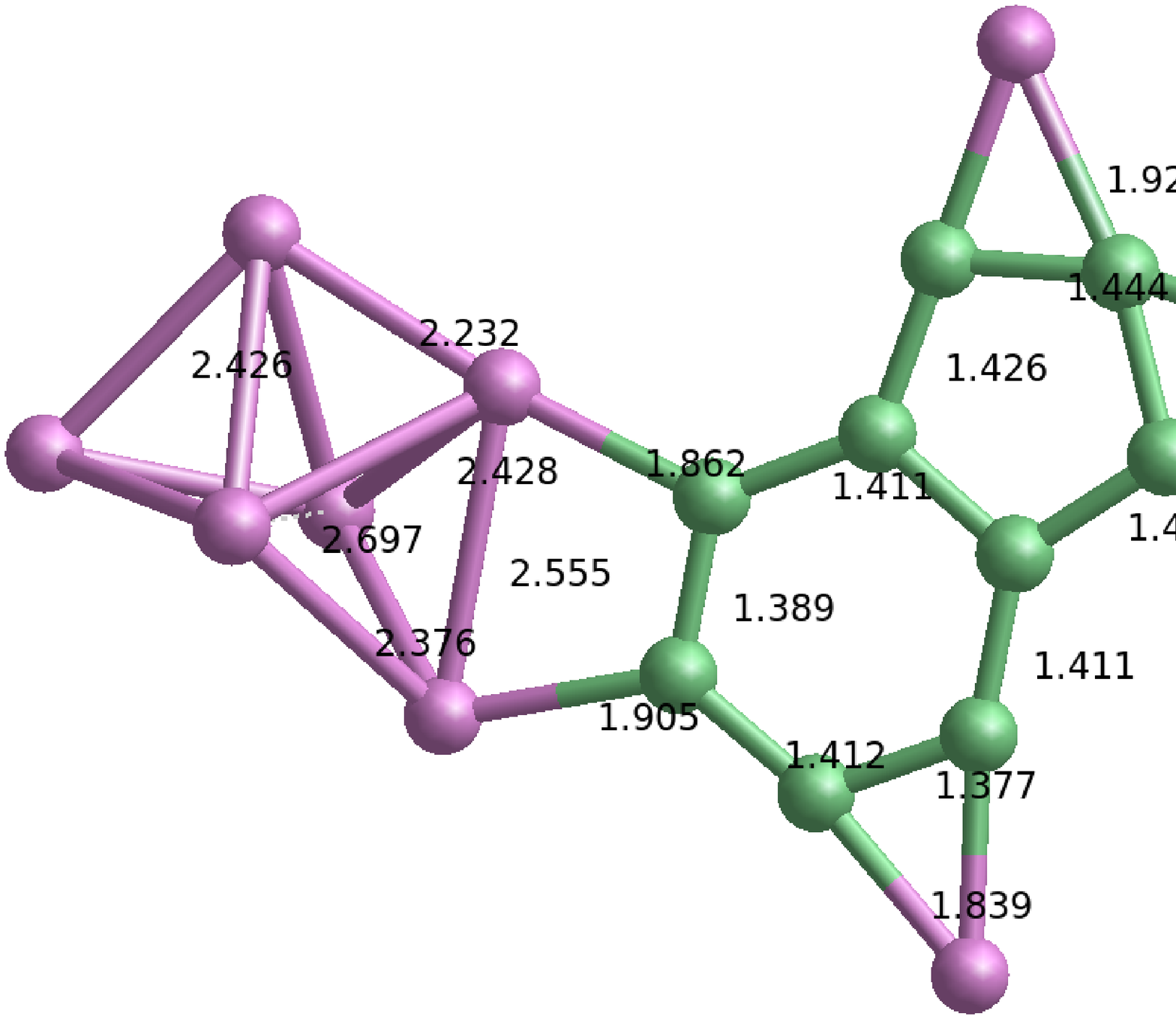}
9
\includegraphics[width=3.5cm]{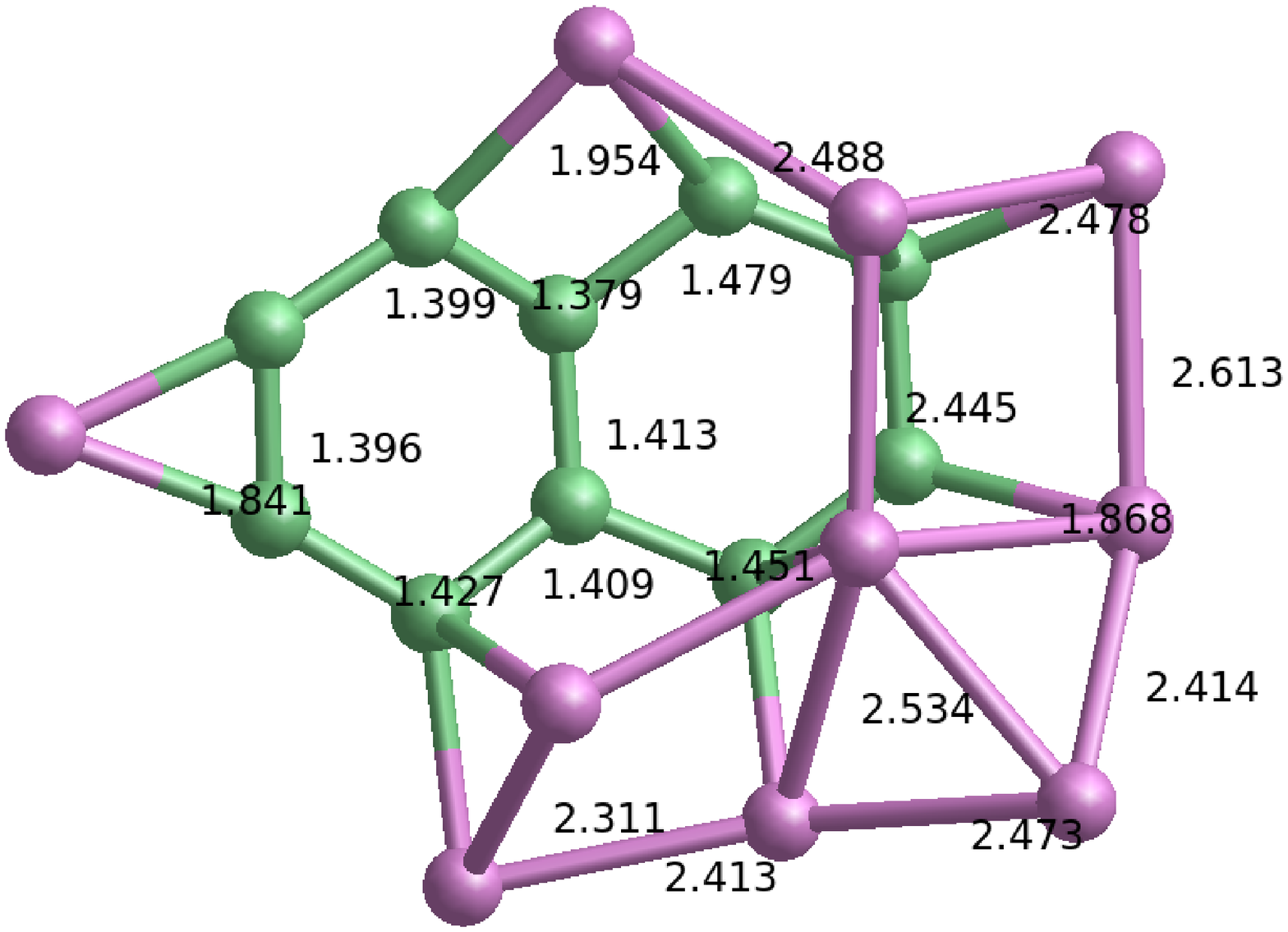}
10
\includegraphics[width=4cm]{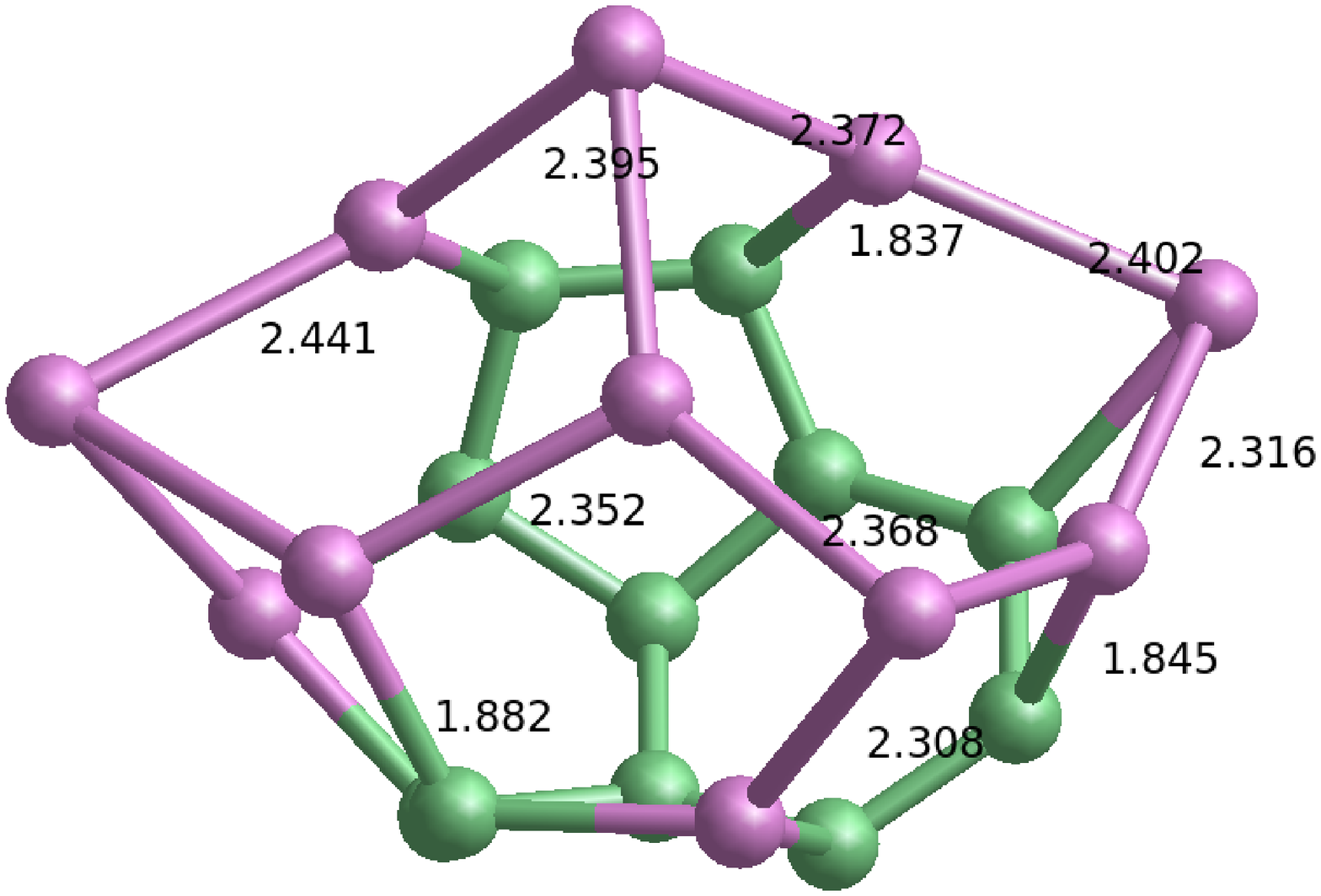}
11
\includegraphics[width=4cm]{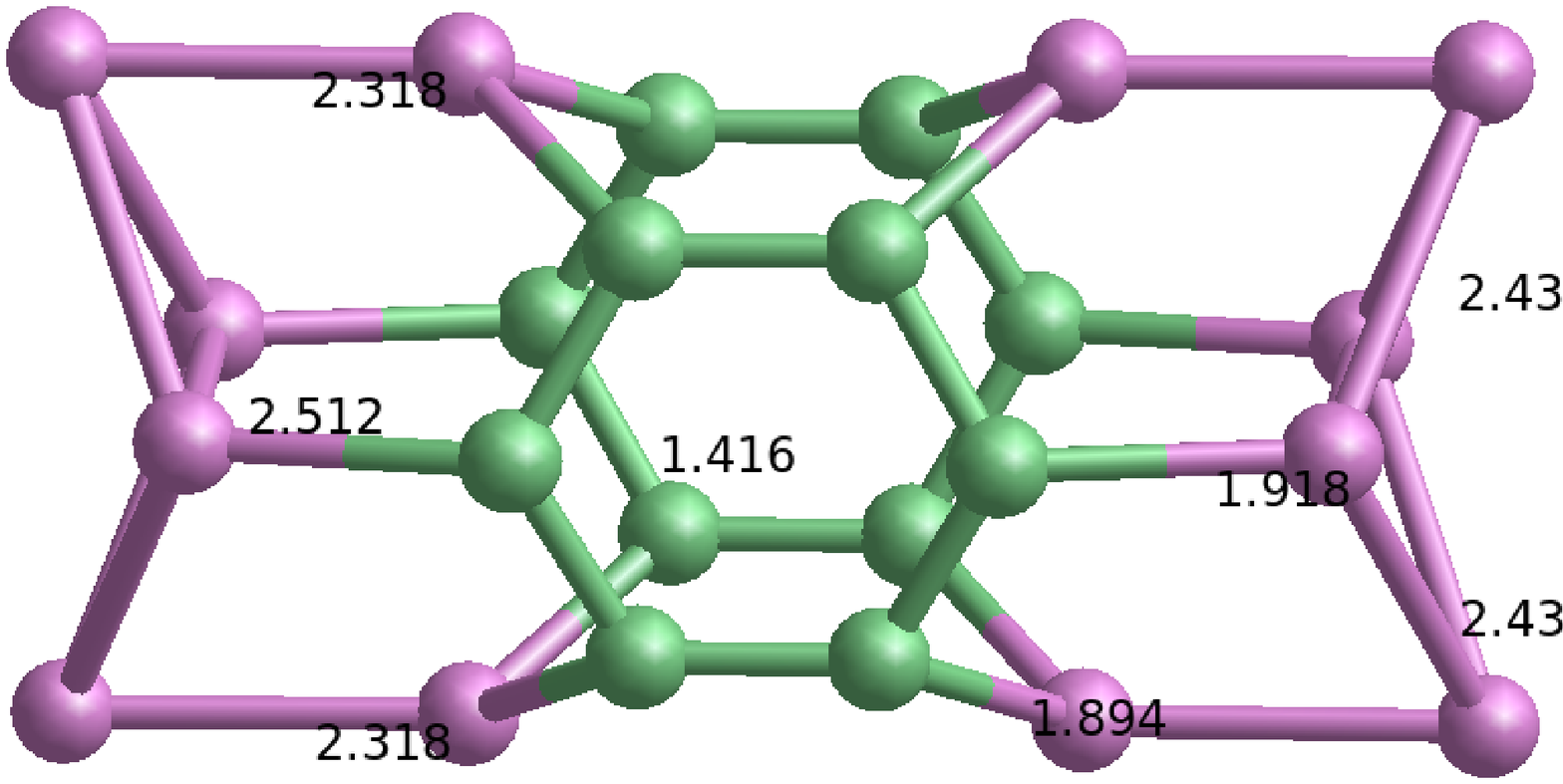}
12
\end{center}
\caption{Electronic structures of the lowest-energy (SiC)\textsubscript{n}\textsuperscript{+}, n=1$-$12, cations. Silicon atoms are displayed in purple, carbon atoms in green. Bond lengths and distances are given in \AA{}\label{fig1}. The cluster size n is given at the right bottom of the corresponding structure.}
\end{figure}

\begin{figure}[h!]
\begin{center}
\includegraphics[width=3.4cm]{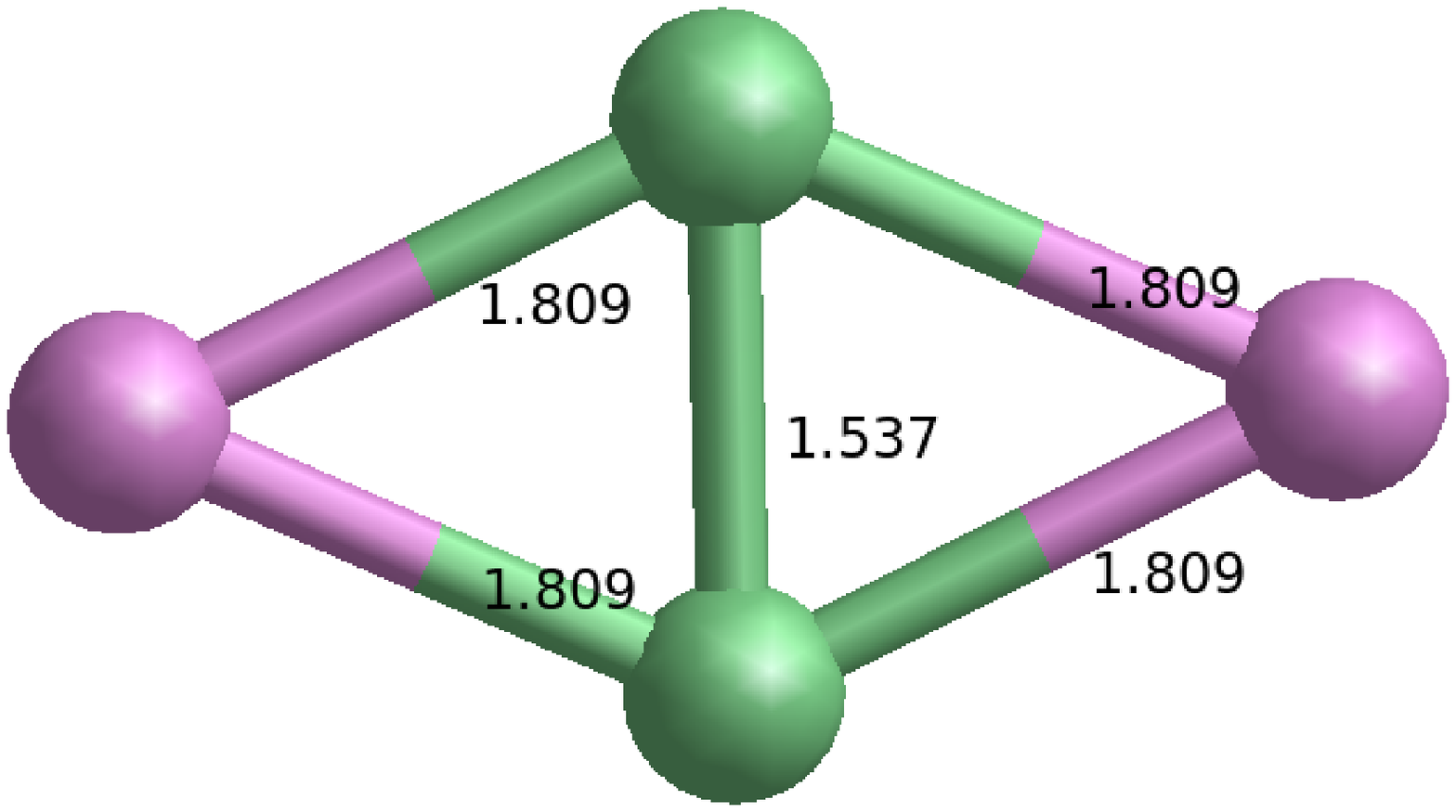}
2 \textit{LM} 1.66 eV
\includegraphics[width=3.4cm]{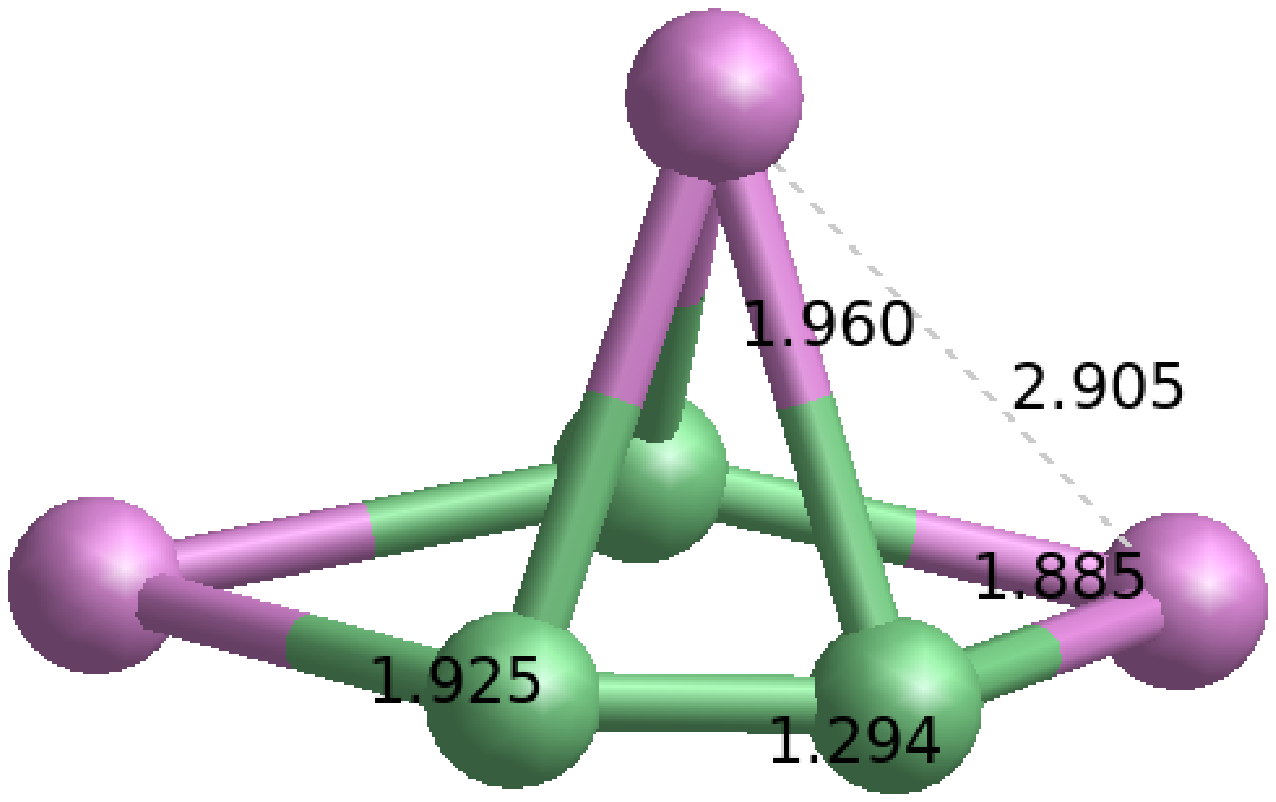}
3 \textit{TS} 1.01 eV
\includegraphics[width=3.4cm]{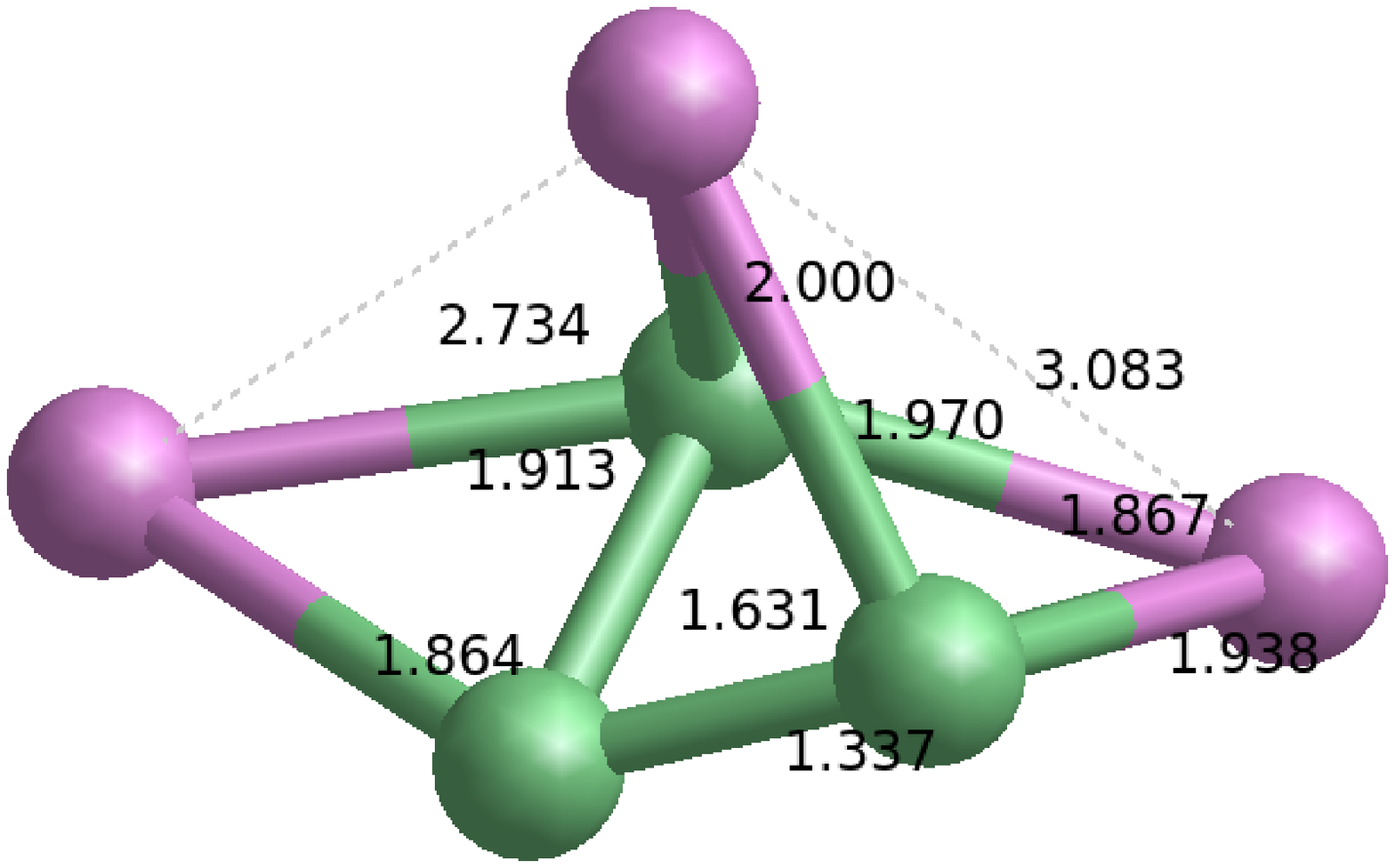}
3 \textit{LM} 0.82 eV
\includegraphics[width=4cm]{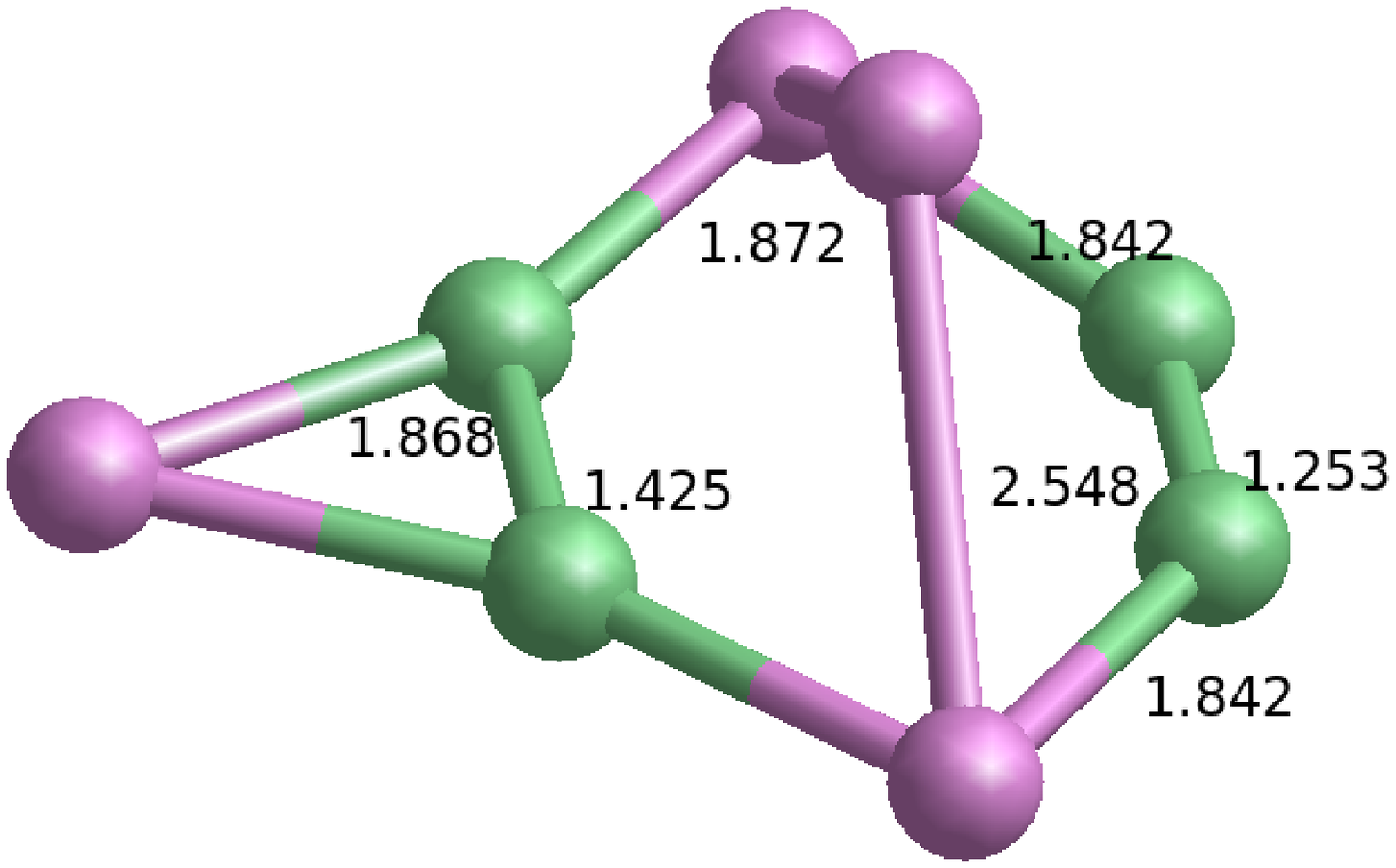}
4 \textit{TS} 1.27 eV 
\includegraphics[width=4cm]{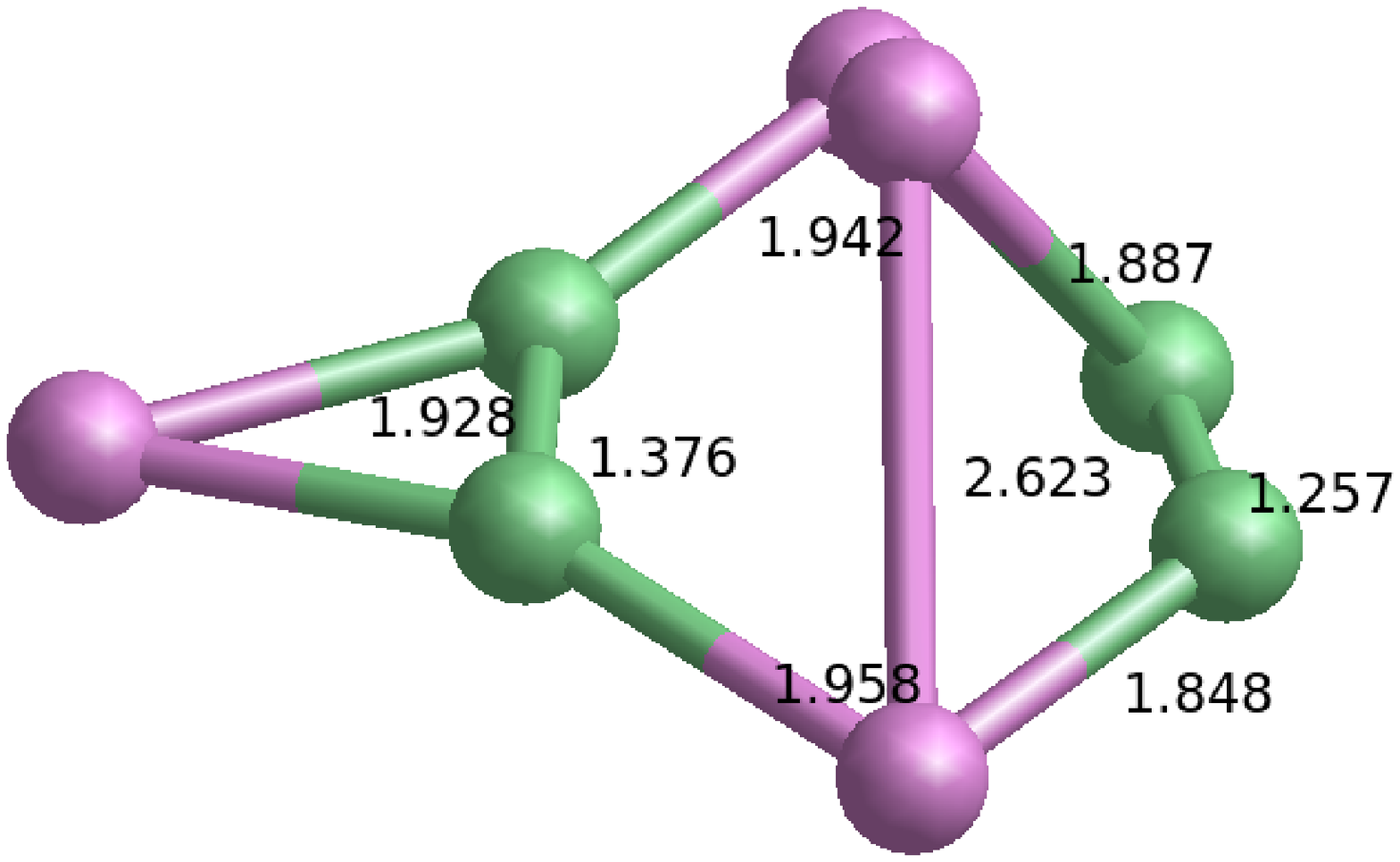}
4 \textit{LM} 0.88 eV
\end{center}
\caption{Electronic structures of metastable (SiC)\textsubscript{n}\textsuperscript{+}, n=2$-$4, cations corresponding to the structural analogues of the neutral Global Minima (GM) clusters. Silicon atoms are displayed in purple, carbon atoms in green. Bond lengths and distances are given in \AA{}. The cluster size n, the type (\textit{LM}: Local Minimum, \textit{TS}: Transition State) and the relative energy is given at the right bottom of the corresponding structure\label{meta}.}
\end{figure}

\begin{figure}[h!]
\begin{center}
\includegraphics[width=2cm]{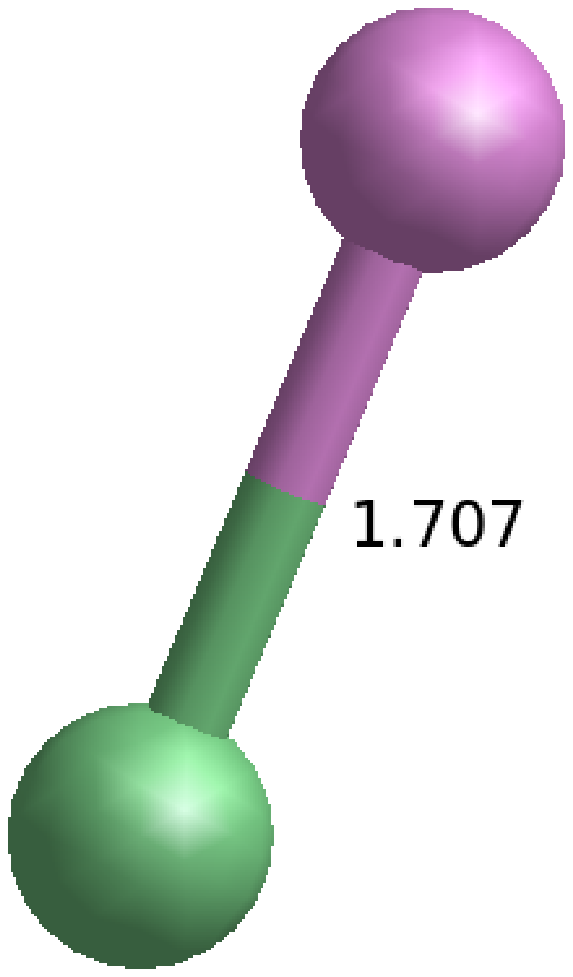}
1
\includegraphics[width=4cm]{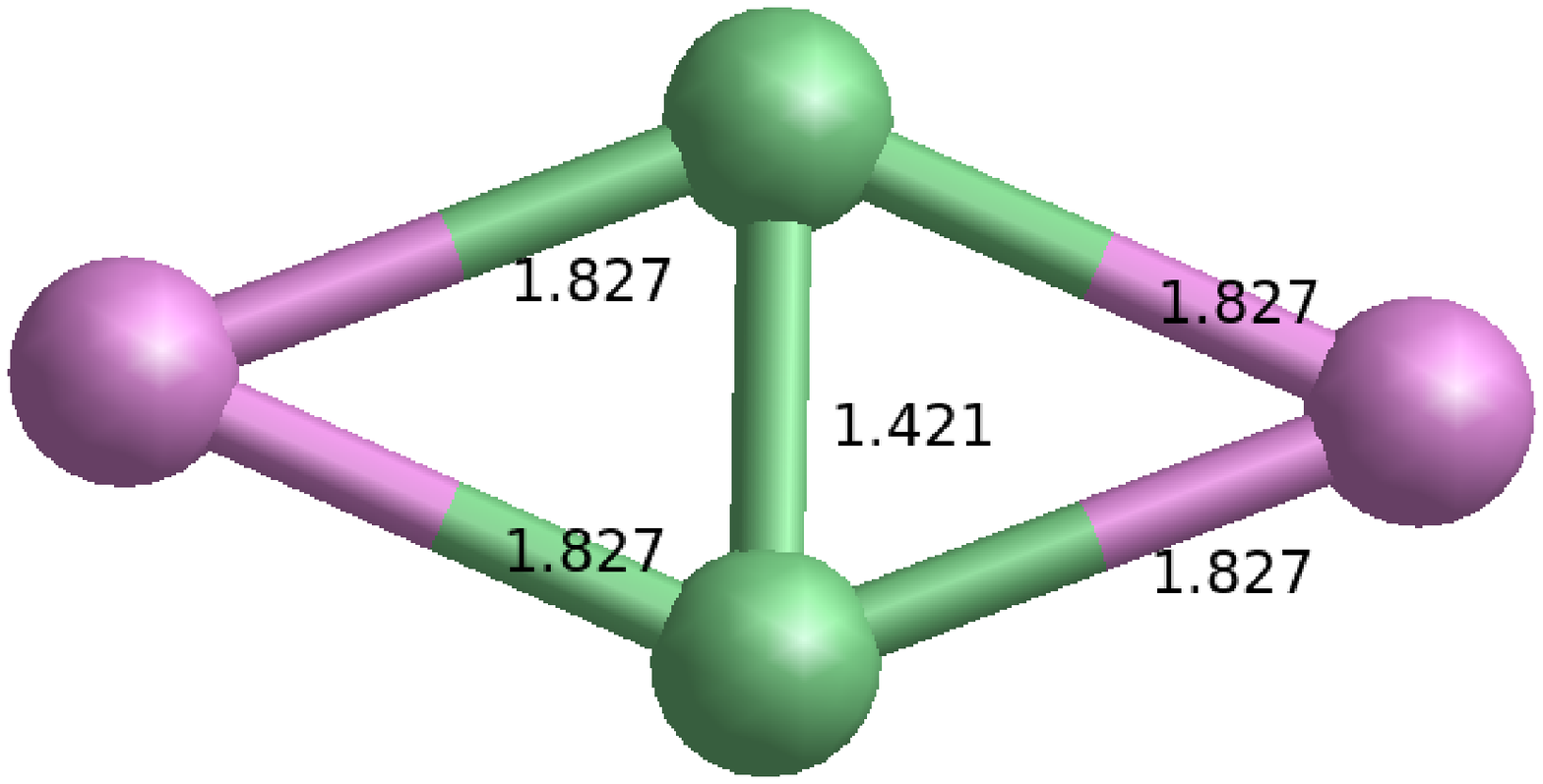}
2
\includegraphics[width=4cm]{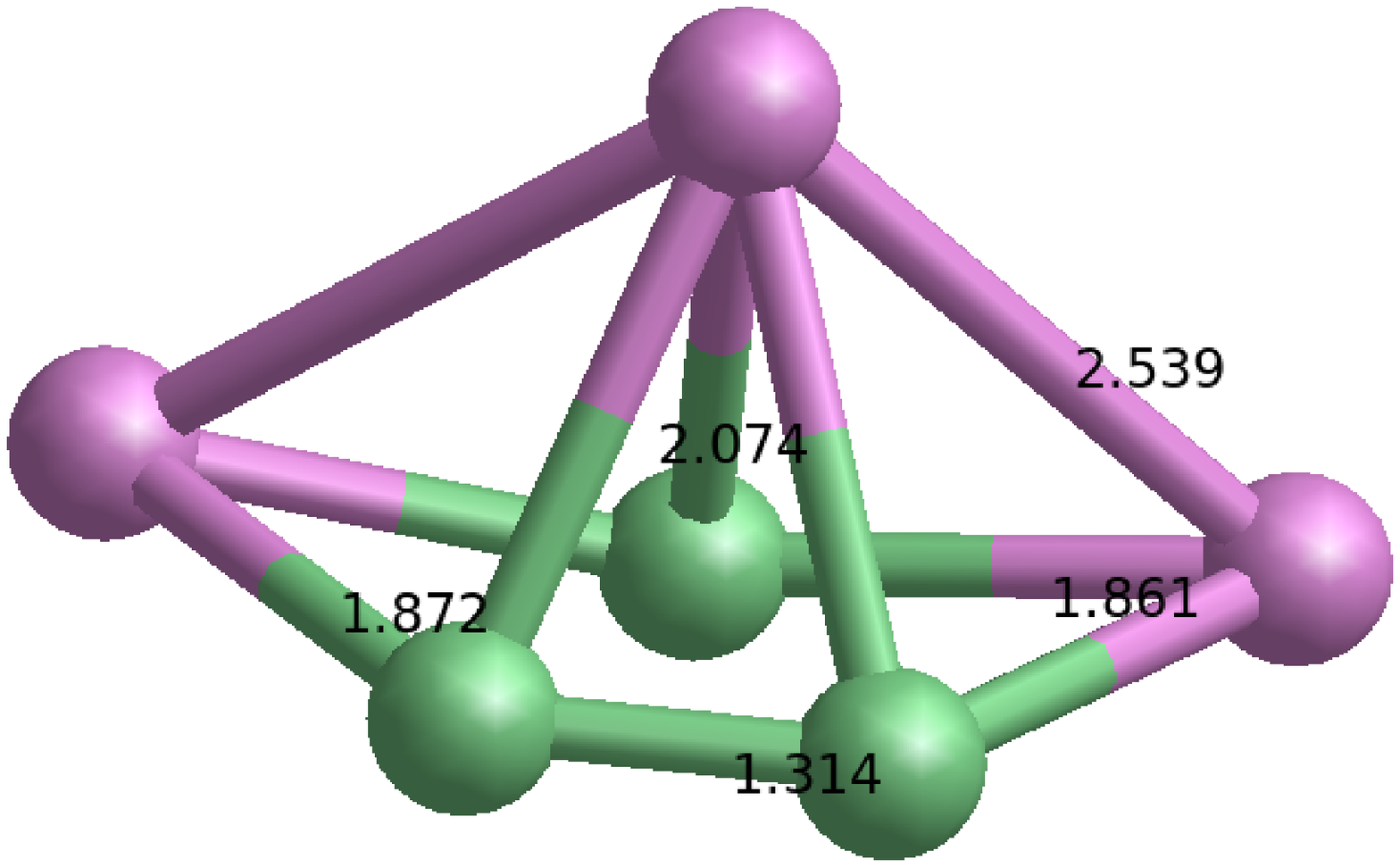}
3
\includegraphics[width=4cm]{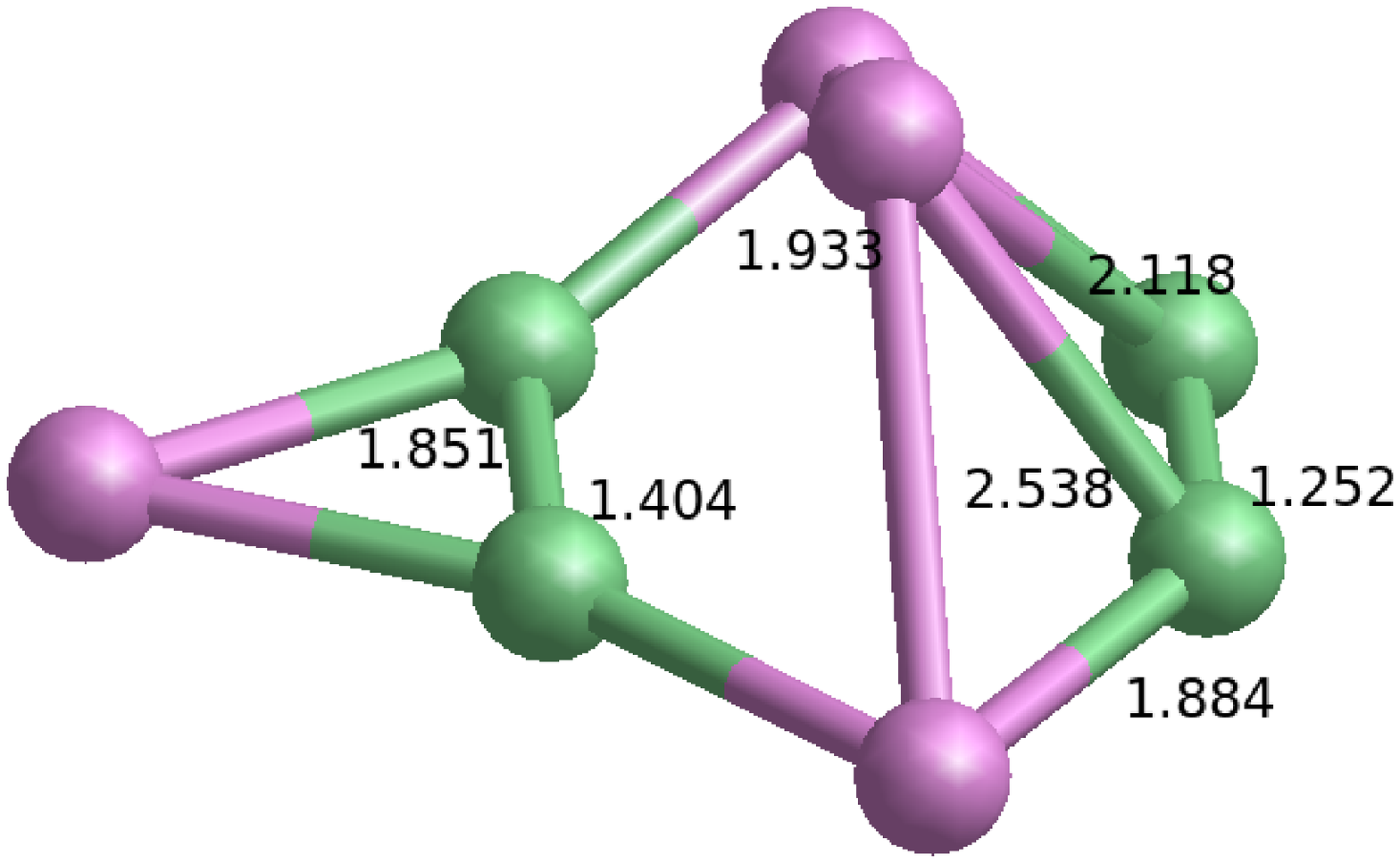}
4
\includegraphics[width=3cm]{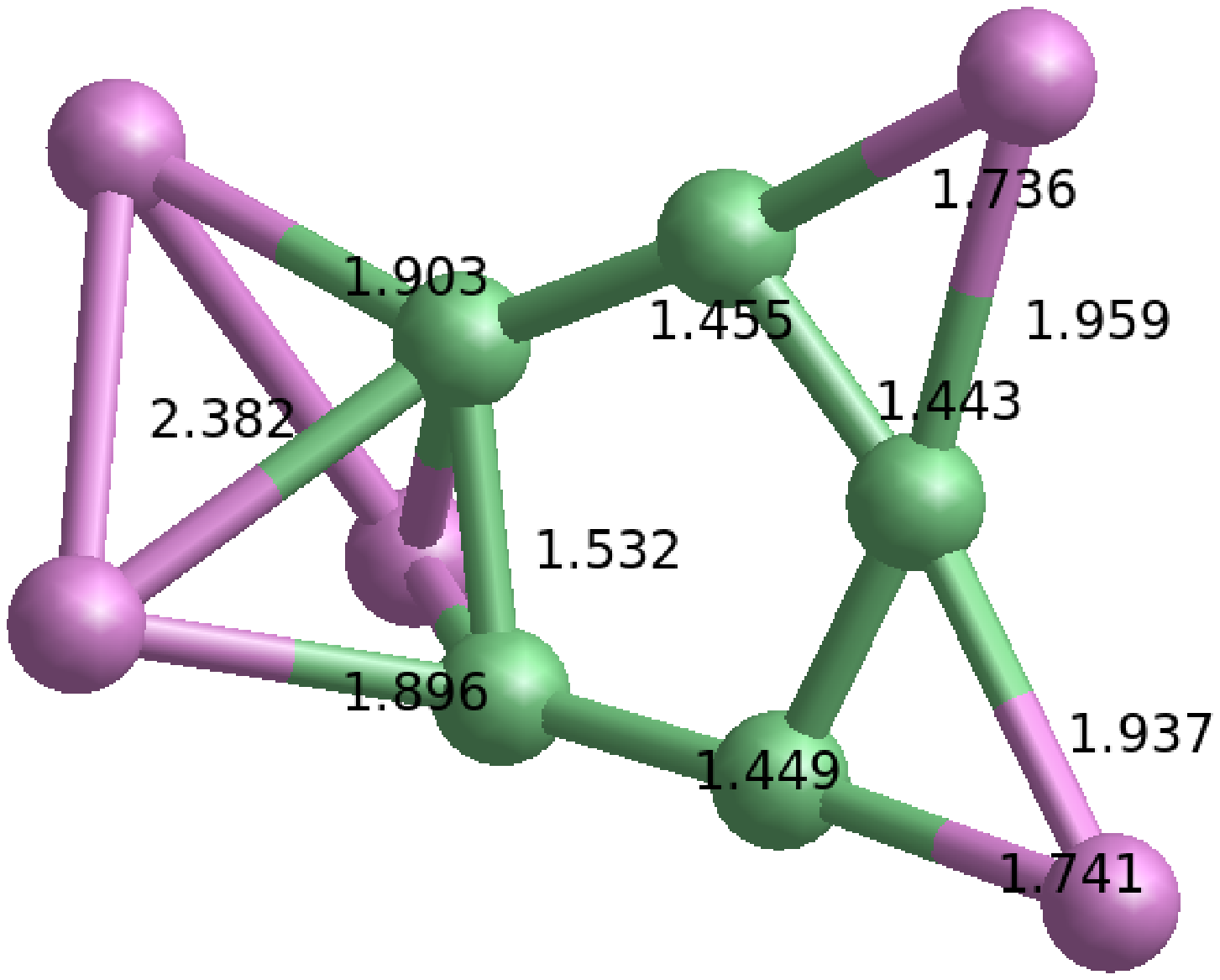}
5
\includegraphics[width=3cm]{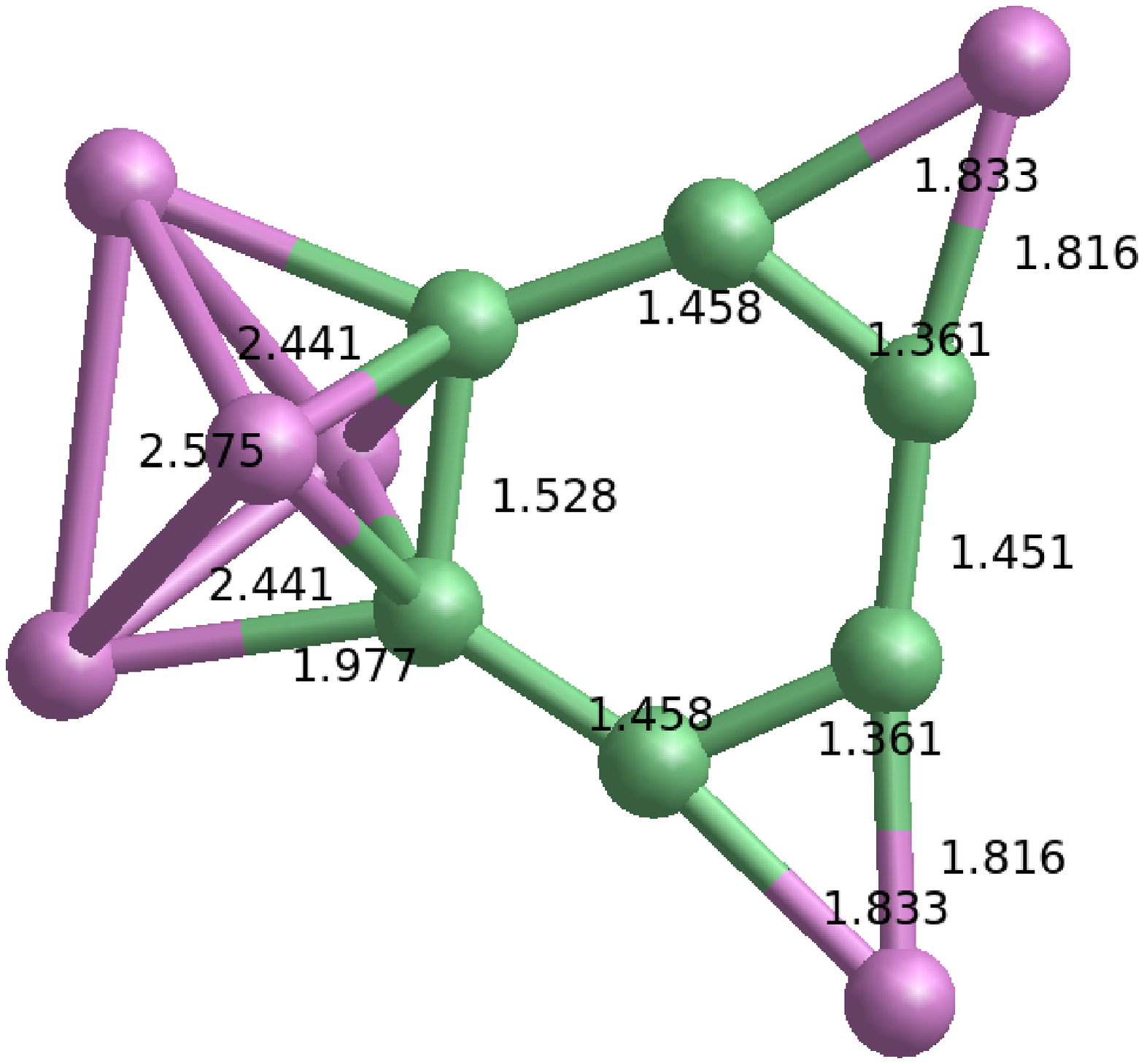}
6
\includegraphics[width=4cm]{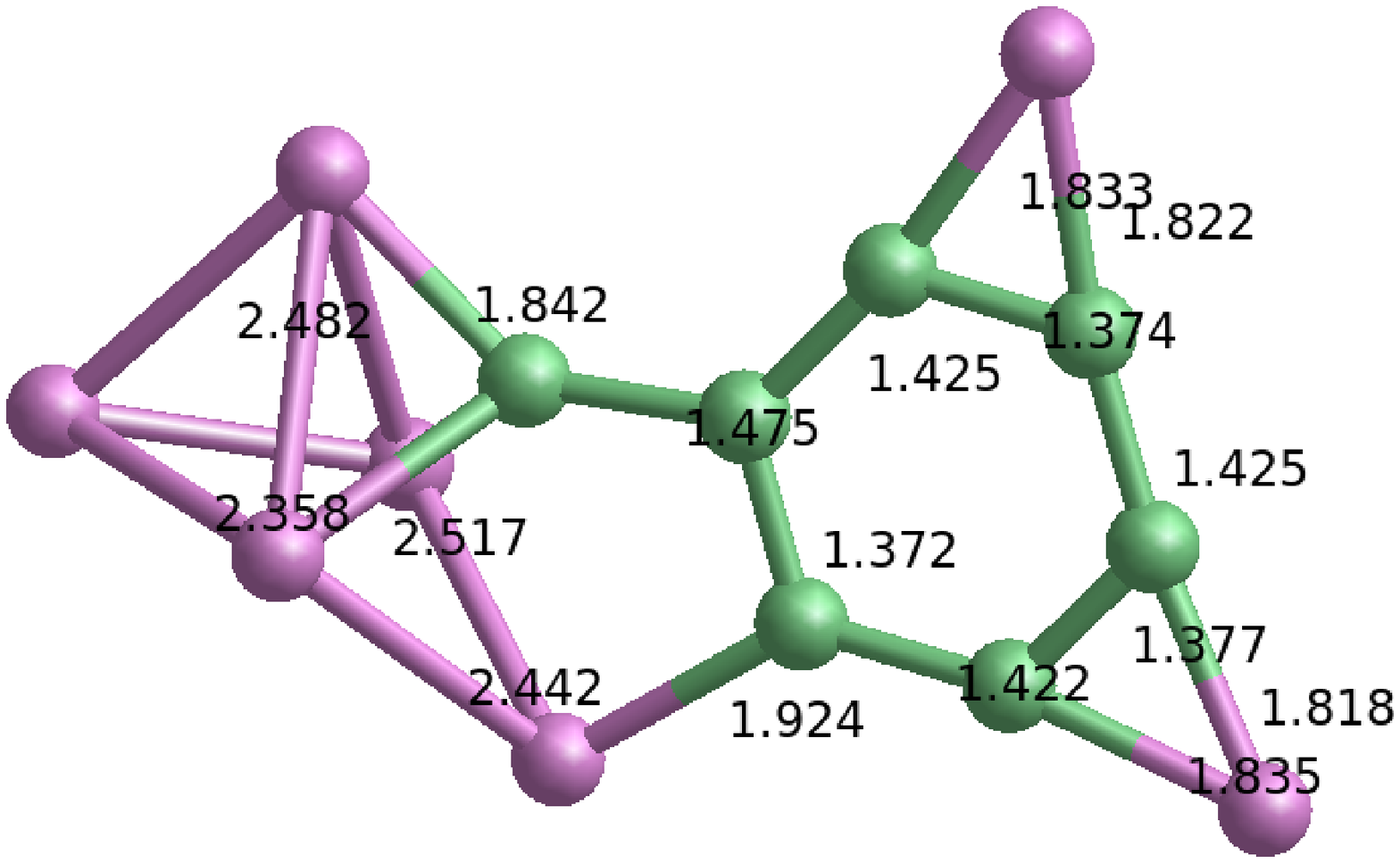}
7
\includegraphics[width=4cm]{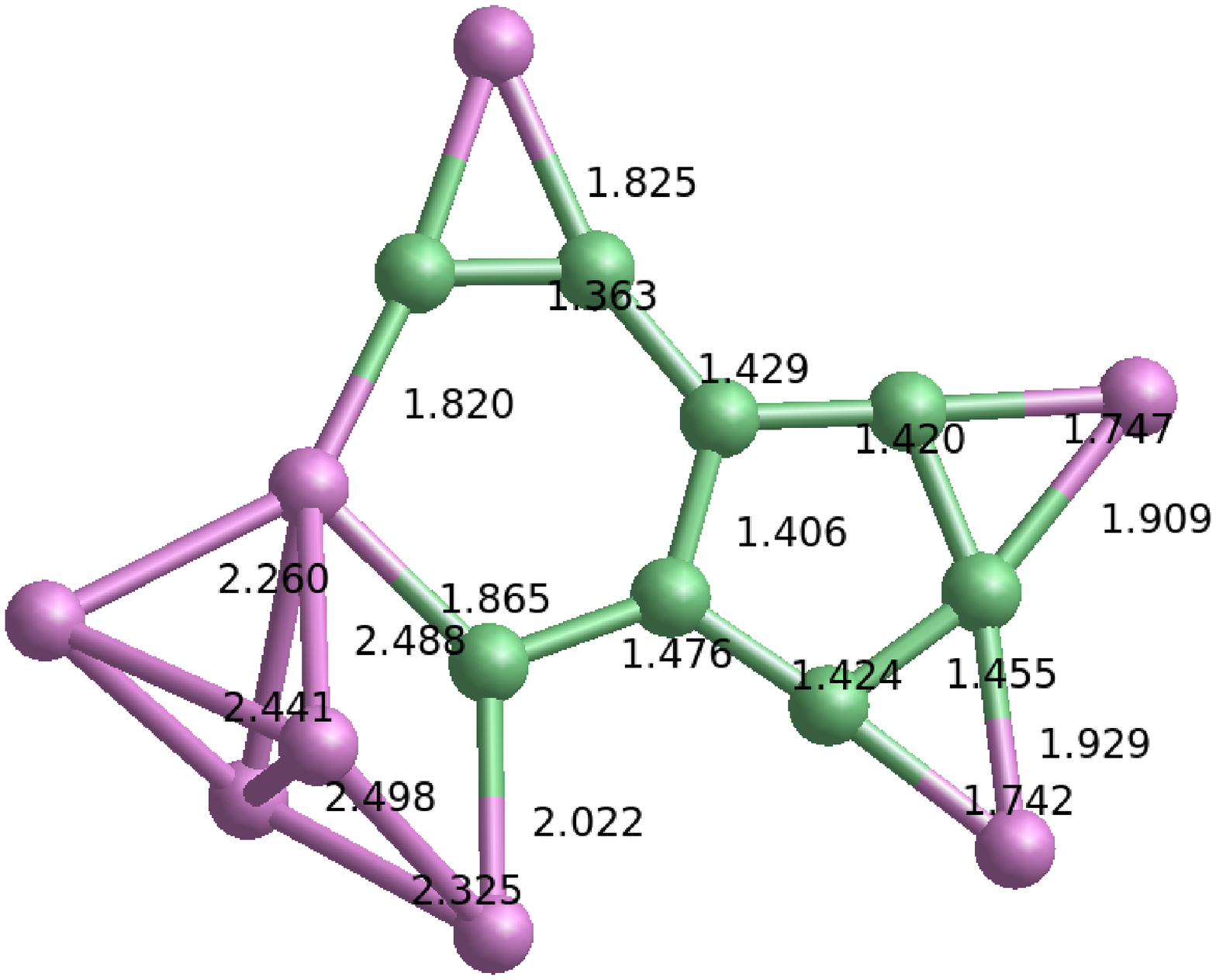}
8
\includegraphics[width=3.5cm]{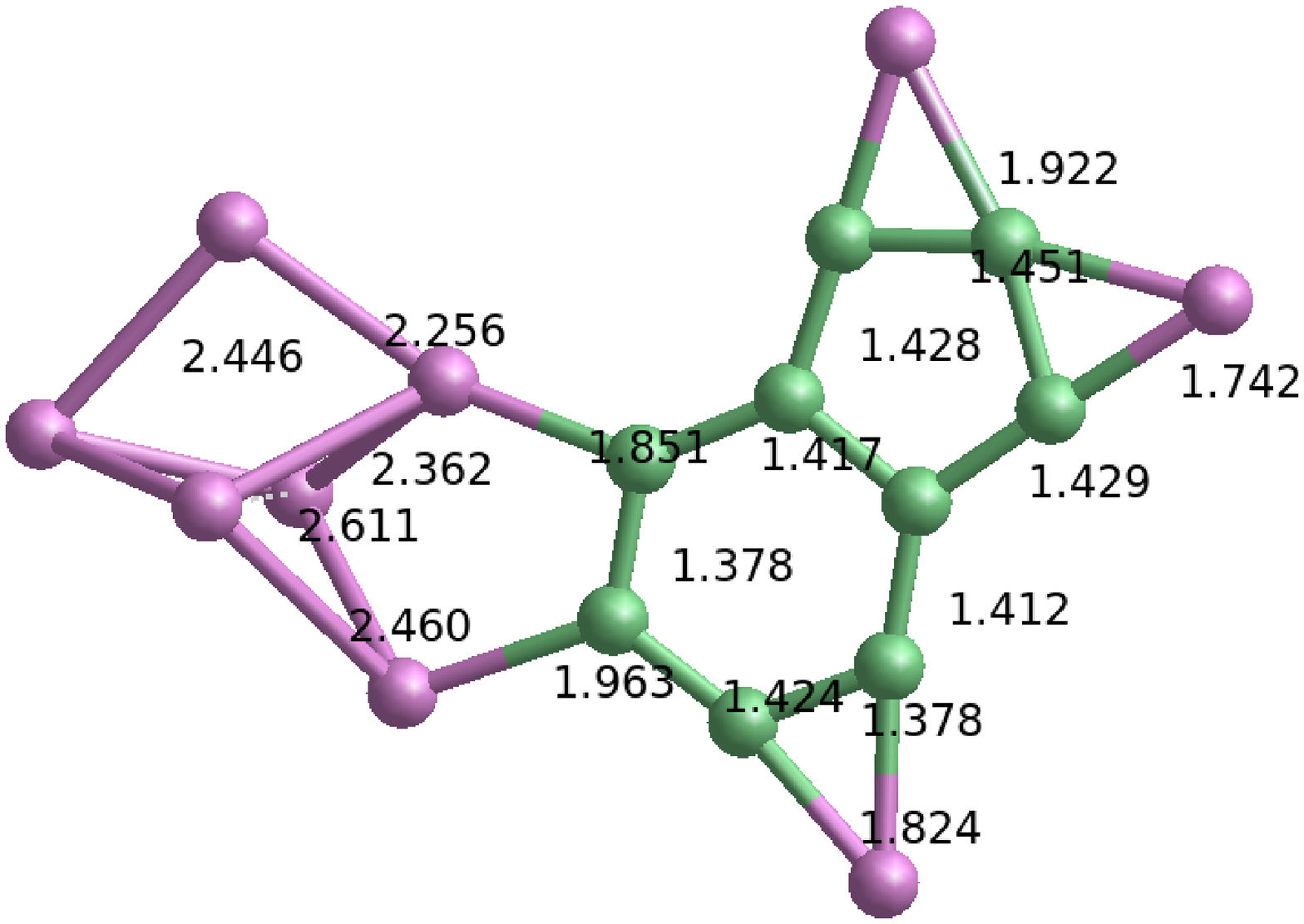}
9
\includegraphics[width=4cm]{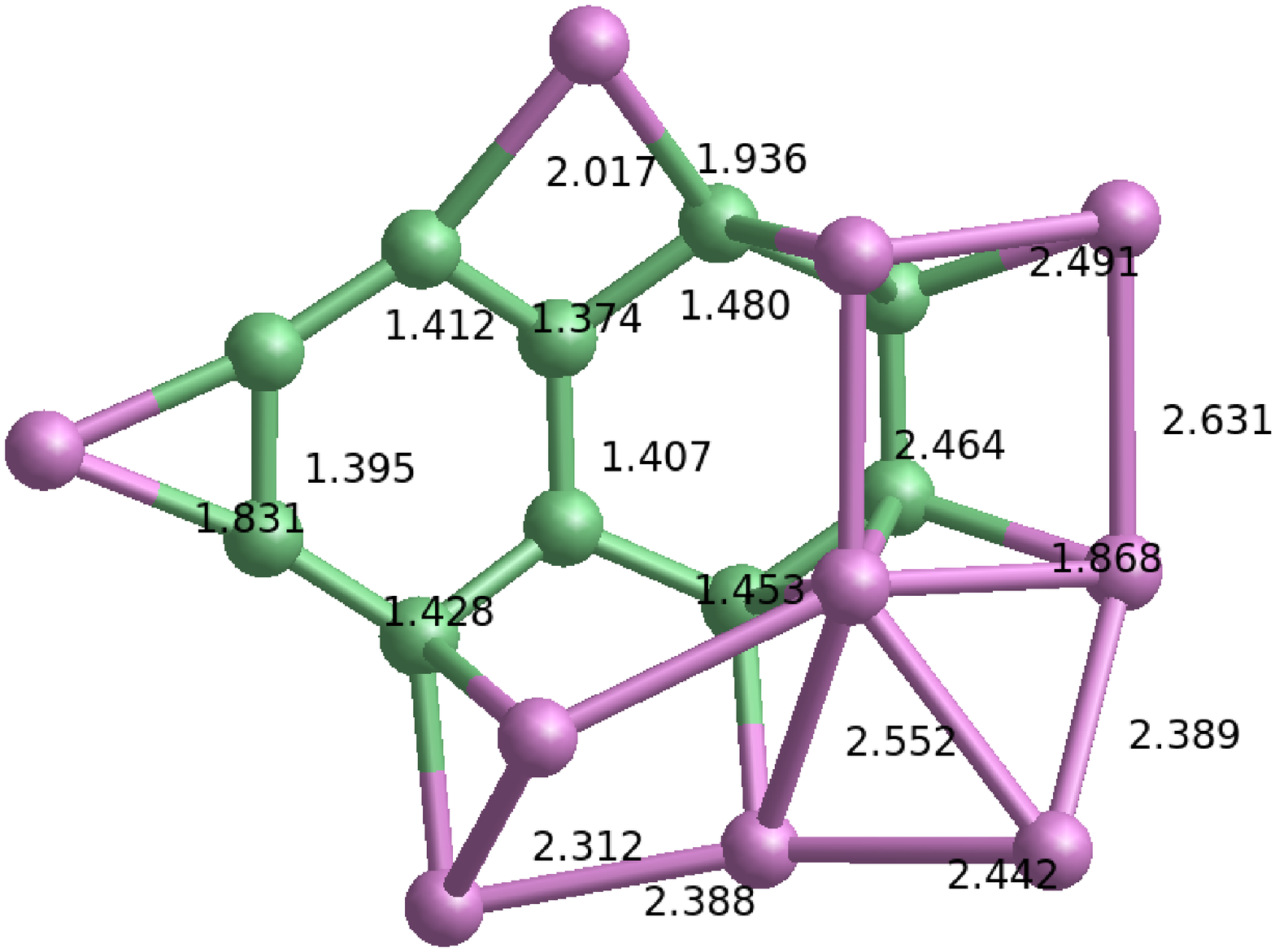}
10
\includegraphics[width=4cm]{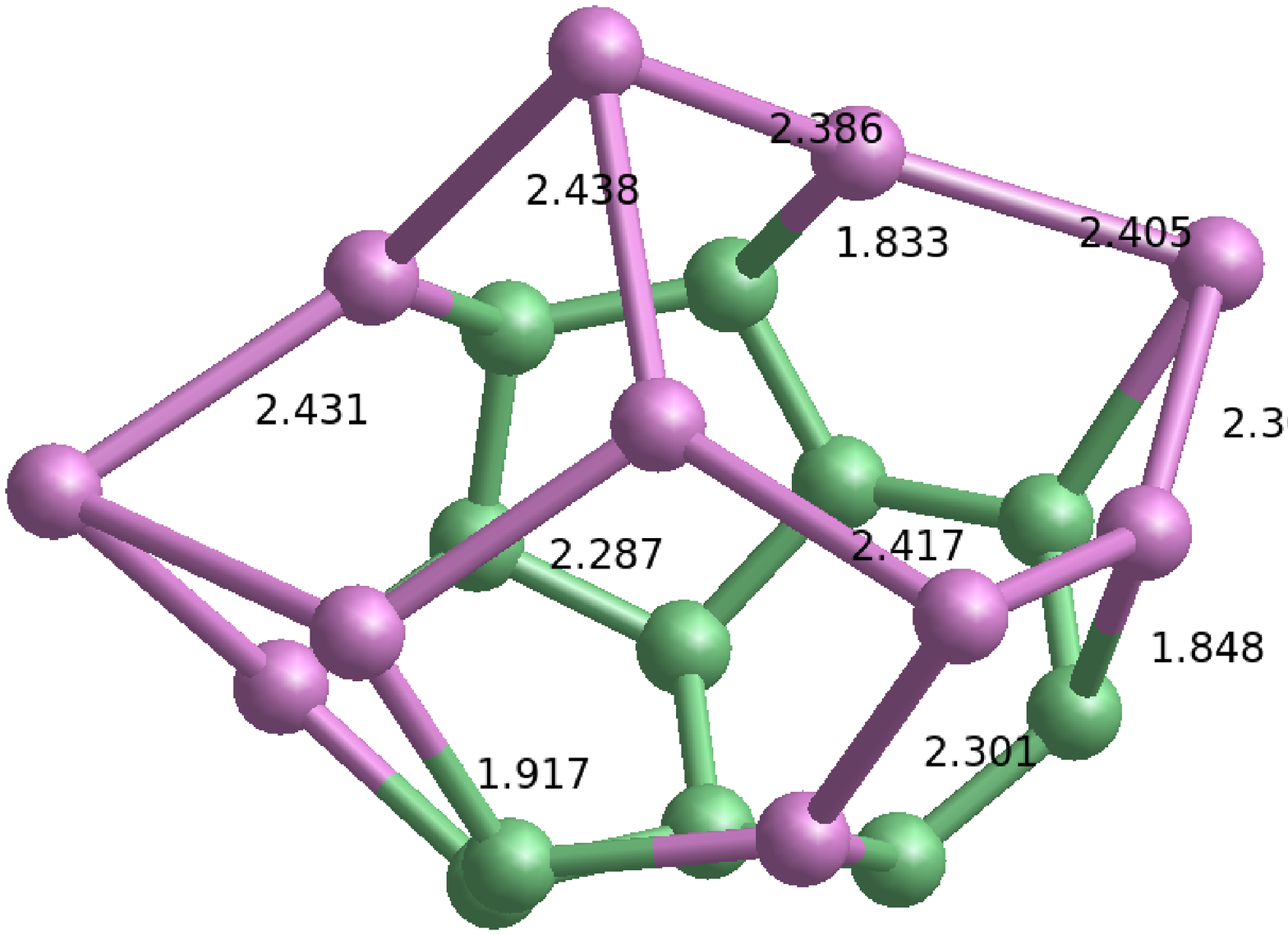}
11
\includegraphics[width=4cm]{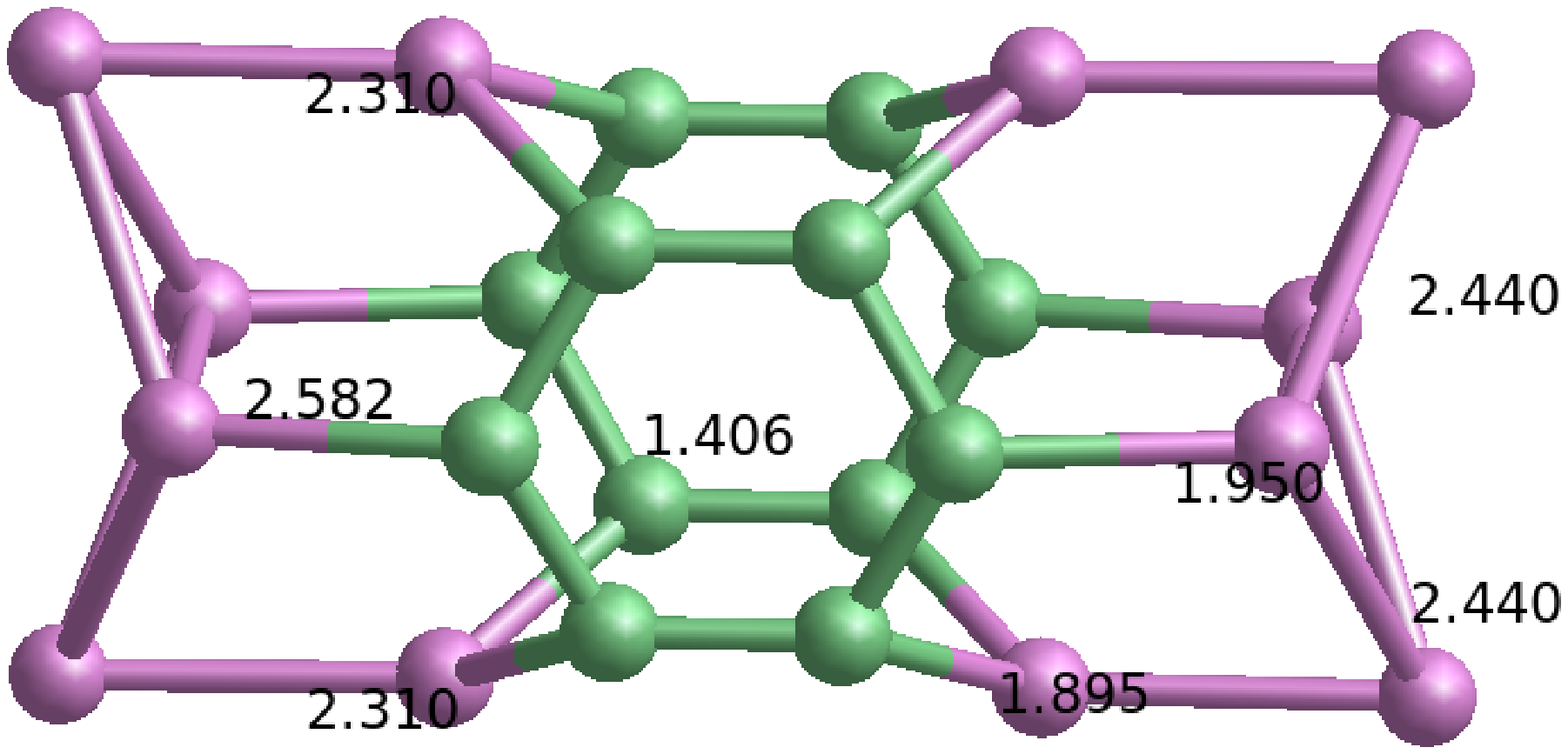}
12
\end{center}
\caption{Electronic structures of neutral (SiC)\textsubscript{n}, n=1$-$12, global minima (GM) candidates. Silicon atoms are displayed in purple, carbon atoms in green. Bond lengths (in \AA{}) are indiacted by numbers. The cluster size n is given at the right bottom of the corresponding GM structure.\label{neutralGM}}
\end{figure}

\begin{figure}[h!]
\begin{center}
\includegraphics[width=3cm]{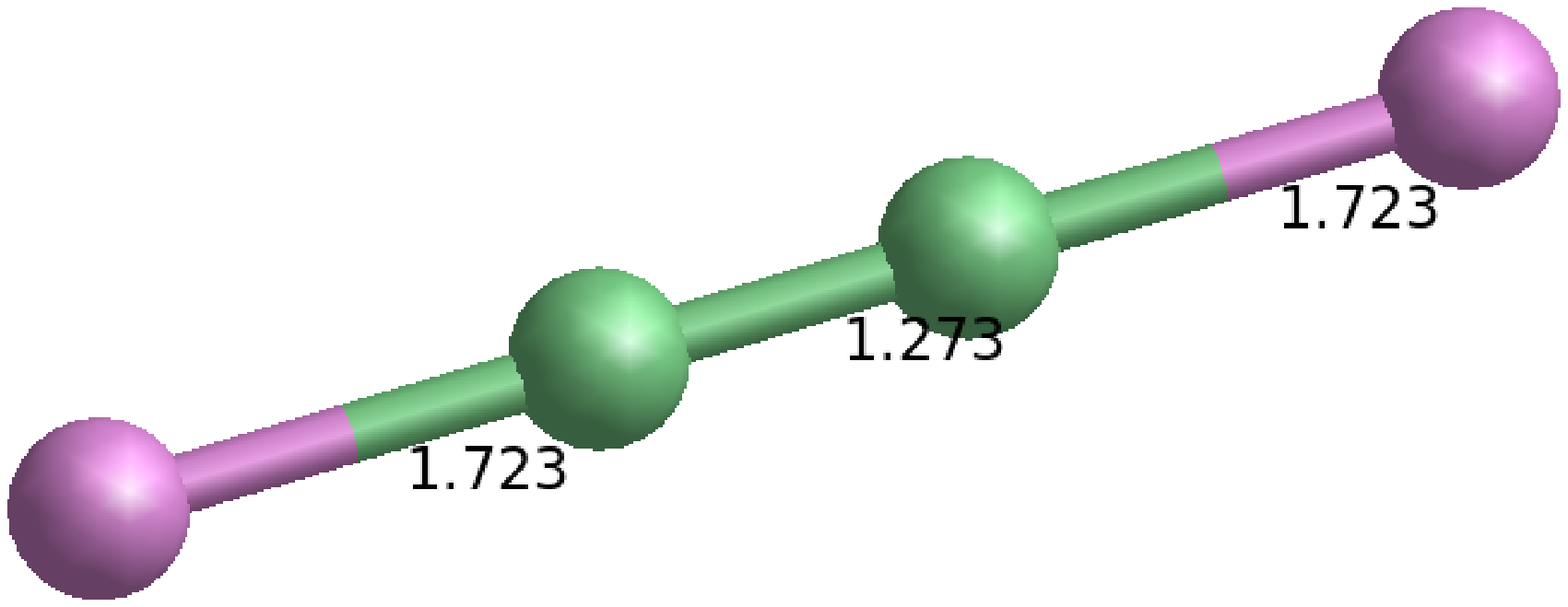}
2 \textit{LM} 0.69 eV
\includegraphics[width=3cm]{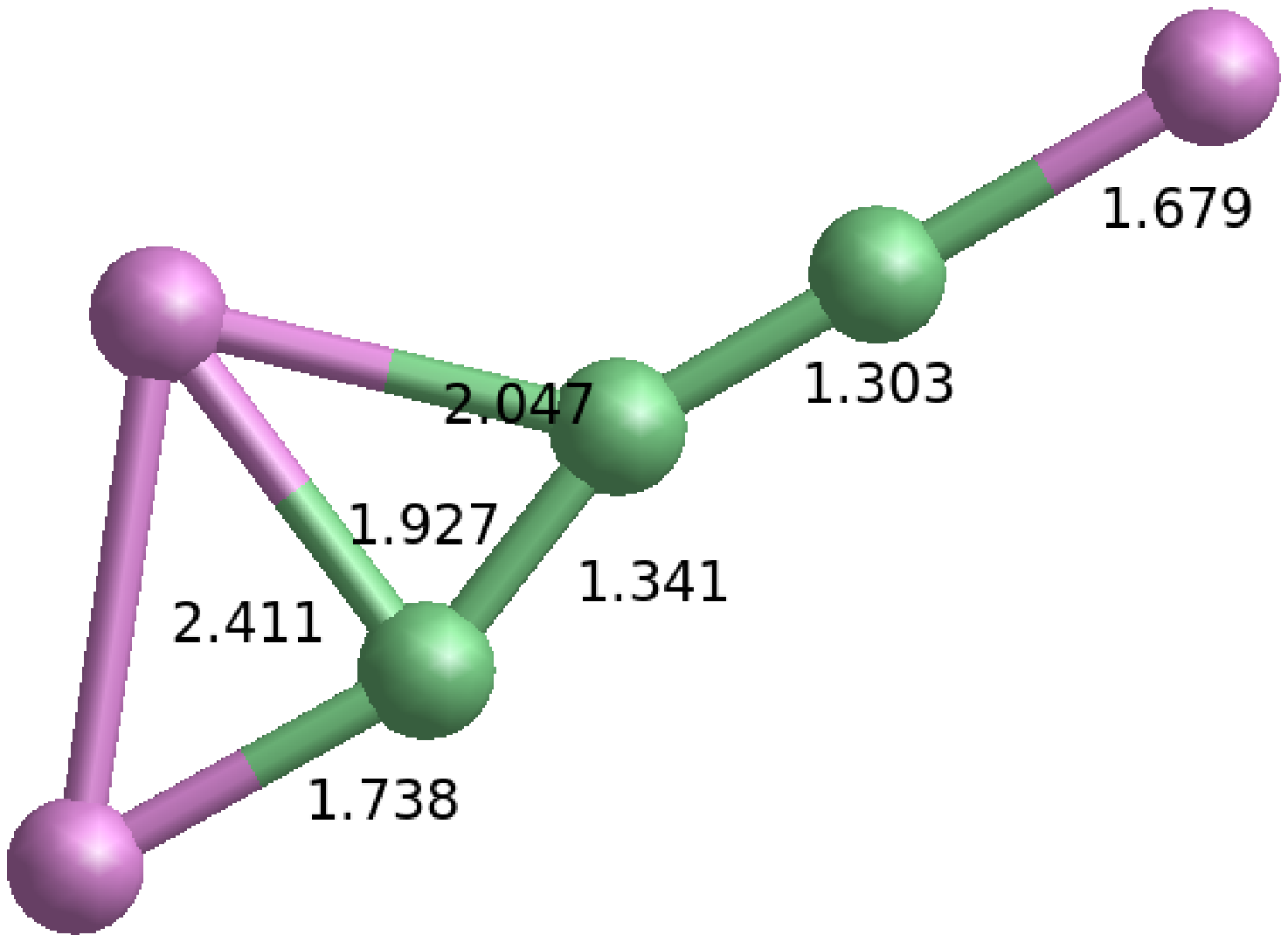}
3 \textit{LM} 0.88 eV
\includegraphics[width=3cm]{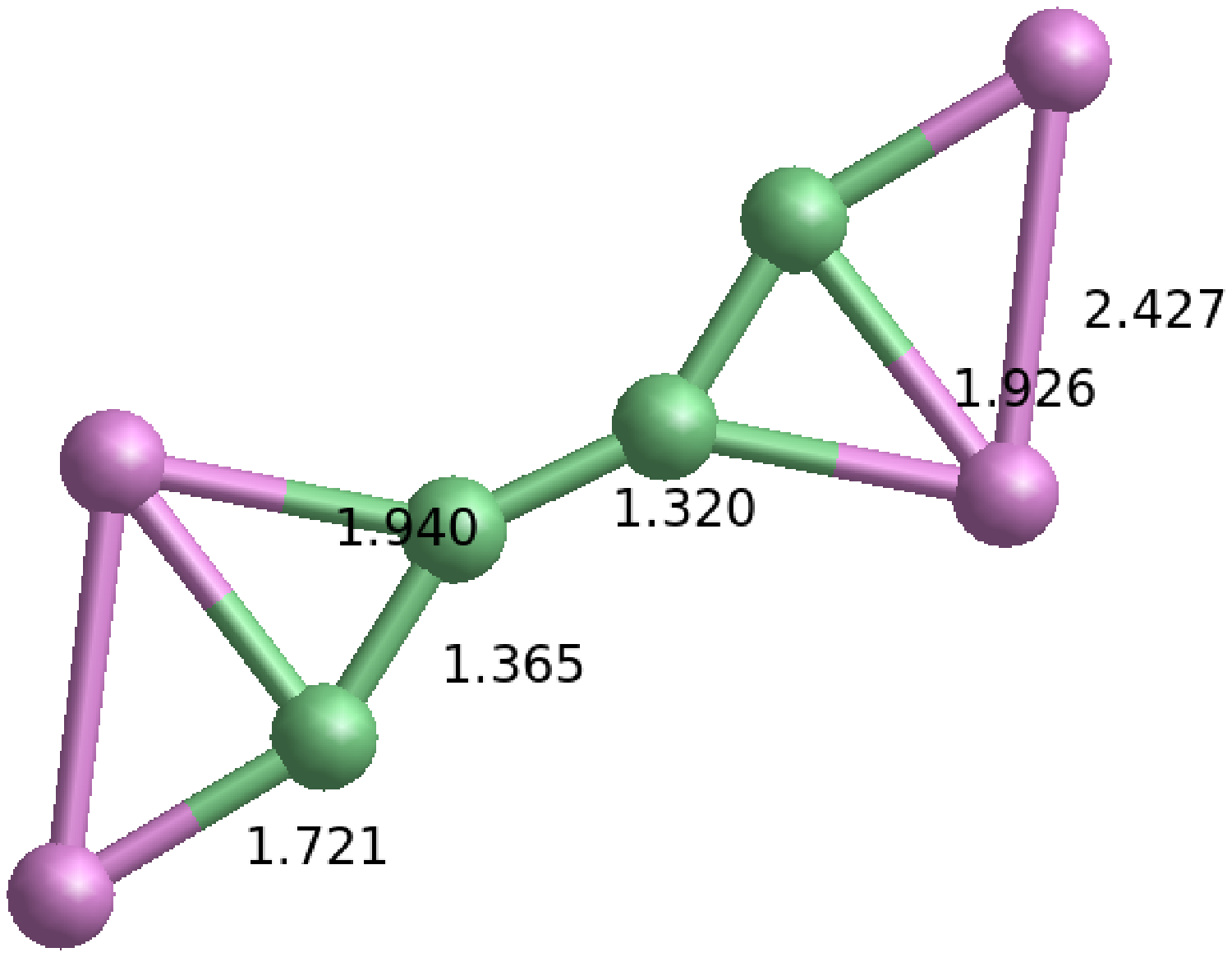}
4 \textit{LM} 0.26 eV
\end{center}
\caption{Electronic structures of neutral (SiC)\textsubscript{n}, n=2$-$4, local minima candidates (denoted as \textit{LM}) corresponding to the structural analogue of the most favourable (SiC)\textsubscript{n}\textsuperscript{+}, n=2$-$4, cations.  The cluster size n at the relative energy is given at the right bottom of the corresponding structure\label{neutralLM}.}
\end{figure}

In this subsection, we discuss the adiabatic ionization 
energies E\textsubscript{n}\textsuperscript{adia} and compare their respective geometries with those of the neutral clusters.
In Figure \ref{fig1} the lowest-energy cation structures are shown. Cation isomers that are derived from neutral GM candidates (i.e. structural twins or analogues), but which are metastable with respect to the lowest-energy cations, are depicted in Figure \ref{meta}. 
For comparison we also display the neutral GM candidates in Figure \ref{neutralGM} and local minima neutral isomers (see Figure \ref{neutralLM}) corresponding to the structural analogues of the lowest-energy cations.\\

The triplet ground state of the neutral SiC molecule has a bond length of 1.707 \AA{} (see Figure \ref{fig1}). The adiabatically optimized SiC\textsuperscript{+} cation is assumed to be a doublet state and has a slightly larger bond length of 1.713 \AA. The energy difference of vertical and adiabatic ionization is negligible (0.002 eV).
In comparison to the experimentally derived ionization energy of 9.2 eV \citep{doi:10.1063/1.1744646} and to the value of 8.7$\pm$0.2 from a previous theoretical investigation \citep{doi:10.1021/j100056a010}, we obtain a higher value (10.64 eV) for the SiC ionization potential.\\
For the SiC dimer, (SiC)\textsubscript{2}, a linear Si-C-C-Si geometry represents the lowest-energy cation structure (see Figure \ref{fig1}), which is different from the diamond shaped neutral GM candidate shown in Figure \ref{neutralGM}.
In comparison with the linear geometry of the neutral dimer (see Figure \ref{neutralLM}), the two Si-C bonds in the cation are larger by 0.055 \AA{} and the C-C bond increases by 0.031 \AA{}.
The adiabatic ionization energy is 7.81 eV with respect to the neutral GM candidate and 7.52 eV with respect to the neutral Si-C-C-Si chain. 
The rhombic (diamond-shaped) (SiC)\textsubscript{2} cation shown in Figure \ref{meta}, resembling the neutral GM candiate, lies 9.87 eV above the latter. In this cation, the C-C bond distance is increased from 1.421 \AA{} to 1.537 \AA{} and the four Si-C bonds are decreased from 1.827 \AA{} to 1.809 \AA{}, when adiabatically ionized. This indicates that a double bonded C=C in the neutral (SiC)\textsubscript{n} becomes single bonded in the related cation.\\
Other isomers show ionization energies in the range of 8.48$-$14.05 eV.         
For n=3, the lowest-energy cation in our set corresponds to a structurally different isomer than the neutral GM (see Figure \ref{fig1} and \ref{neutralGM}). It has an energy of 7.94 eV above the neutral GM and 7.04 eV above its neutral structural twin. The largest geometric change upon the loss of an electron consists in the breaking of a bond (marked with dashed line and labelled with a distance of 2.953 \AA{} in Figure \ref{fig1}), which is present in the neutral isomer (2.047 \AA{}). 
The other bond lengths change by 0.03$-$0.11 \AA{}. 
The cationic twin of the neutral GM displayed in Figure \ref{meta} has an energy 1.01 eV above the lowest-energy cation and is a TS, whose distorted, asymetric geometry results in a real minimum with a relative energy of 0.88 eV. 
These cations show increased Si-Si distances as compared with the neutral GM. 
Further investigated trimer isomers have ionization energies in the range of 8.20$-$10.26 eV.\\ 
The adiabatic ionization of different (SiC)\textsubscript{4} isomers shows that the lowest-energy cation is a structural analogue of the second most favourable neutral structure (see Figure \ref{fig1} and \ref{neutralLM}). The GM cation has an adiabatic ionization energy of 7.36 eV with respect to the neutral GM and 7.07 eV with repect to the neutral structure analogue.
It consists of a C\textsubscript{4} chain in a trans configuration with two Si atoms arranged at each of the two tails of the chain. Upon the loss of an electron, most significantly, the two Si-Si bonds increase by 0.13 \AA{} among minor changes in the other bond lengths.  
As for n=3, the optimized (SiC)\textsubscript{4}\textsuperscript{+} structural twin of the neutral GM shows an imaginary vibrational frequency and is a TS with an energy of 8.63 eV above the neutral GM.
To obtain a real minimum, a further optimization is performed leading to a symmetry breaking of the C\textsubscript{2}-Si-C\textsubscript{2}-Si ring structure by increasing two Si-C bond lengths and by tilting the quasi planar ring structure. The local minimum lies about 0.40 eV below the TS. Other cation isomers show ionization energies of 7.41$-$7.84 eV \\ 
The most favourable (SiC)\textsubscript{5}\textsuperscript{+} cation resembles its neutral analogue and has an ionization energy of 7.61 eV. However, in contrast to its neutral counterpart the two out-of-plane Si atoms increase their distance to each other by 0.25 \AA{}. 
This structure is a TS and a subsequent optimization to a real minimum leads (as for n=3 and 4) to a breaking of the C\textsubscript{2v} symmetry. In the real minimum cation, the two out-of-plane Si atoms have distance of 2.63 \AA{} to each other (like in the neutral cluster), but are shifted into the display plane, which reduces the energy by 0.14 eV (as compared to the TS).
The ionization energies of other investigated isomers range from 7.90 eV to 8.32 eV\\    
For n=6, the lowest-energy cation is a structural analogue of the neutral GM candidate with an ionization energy of 7.46 eV.
Its overall geometry hardly changes by the adiabatic ionization. The C-C and Si-C bond distances are largely preserved. The biggest change occurs at two Si atoms increasing their distance by 0.30 \AA{}. This likely to be the place, where the loss of the electron occured during the adiabatic ionization, which accounts for 0.44 eV. We find ionization energies of 7.50$-$8.45 eV for other hexamer cation isomers.\\  
The lowest-energy (SiC)\textsubscript{7}\textsuperscript{+} cation resembles the neutral (SiC)\textsubscript{7} GM cluster and lies 7.25 eV above it. The largest changes in the bond distances occur at Si-Si bonds that lie out-of-plane. The difference in the vertical and adiabatic ionization energy (0.31 eV) is smaller than for n=6 and n=8. Further (SiC)\textsubscript{7}\textsuperscript{+} isomers exhibit ionization energies of 7.60$-$8.26 eV.\\  
For n=8, an ionization energy of 7.25 eV is found for the lowest-energy cation, which is a look-alike of the neutral GM candidate.
The largest atomic mobility during the adiabatic ionization is accomplished by the out-of-plane Si atoms.
These two Si atoms have moved closer to the C atom and are now 4-fold coordinated, in comparison to the neutral (SiC)\textsubscript{8} cluster, where these Si atoms are 3-fold coordinated. The adiabatic structure change results in a lower energy of 0.37 eV. For other isomers we find ionization energies of 7.62$-$7.73 eV.\\
For n=9, the neutral GM and lowest-energy cation exhibit a similar shape and differ by 6.96 eV in energy. 
Its adiabatic ionization affects predominantly the silicon-segregated part of the isomer. The two Si atoms out-of-plane increase their distance to each other, whereas two Si atoms in the plane move closer to each other accounting for 0.38 eV. Other local minima have ionization energies of 7.11$-$7.64 eV.\\  
For n=10, both the neutral (SiC)\textsubscript{10} clusters and its cation (SiC)\textsubscript{10}\textsuperscript{+}, do not exhibit planar carbon ring structures, but are strained. This is the smallest cluster / cation size, where the C atoms do not lie in a planar configuration. 
The lowest-energy (SiC)\textsubscript{9}\textsuperscript{+} cation geometry corresponds to the one of the neutral GM candidate and has a relative energy of 6.95 eV.
The largest impact on the geometry is found to be located at a single Si atom. In the cation this Si atom is connected to the silicon subcluster, whereas in the neutral cluster it binds only to C atoms. This Si migration causes an energy difference of 0.25 eV. Other cation isomers show ionization energies of 7.30$-$7.84 eV.\\  
For the neutral (SiC)\textsubscript{11} GM candidate, the adiabatic optimization of the corresponding cation failed. However, we found a very similar neutral (SiC)\textsubscript{11} cluster with a relative energy just 0.10 eV above the GM candidate, for which an adiabatic ionization optimization could be achieved.
With an ionization energy of 6.64 eV it also corresponds to the lowest-energy cation configuration for n=11.
By inspecting the bond lengths of the neutral and cationic (SiC)\textsubscript{11} structure, 
we find that none of them changes significantly (max of 0.07 \AA{} / bond) during the adiabatic ionization, which is reflected in the comparatively low energy difference of 0.18 eV. Higher-lying cation isomers have energies 6.92$-$7.35 eV above the neutral GM.\\
The GM candidate of neutral (SiC)\textsubscript{12} is found to be a bucky-like, spherical structure with strictly alternating Si-C bonds. However, our calculations for vertical and adiabatic ionization energies did not converge. 
The ionization of fullerene-shaped clusters is a dynamic and complex problem and requires high-level calculations and analyses \citep{D0CP01210F}. For example, Buckminster fullerenes cations (C\textsubscript{60}\textsuperscript{+}) undergo dynamic and pseudo Jahn-Teller effects lowering the icosahedral symmetry upon (photo-)ionization. Owing to these challenges, that are beyond the scope of this paper, we omit an in-depth investigation of the ionization of the ``bucky'' (SiC)\textsubscript{12} GM candiate isomer. 
Therefore, we calculate the ionization energies of an atom-segregated symmetric “butterfly”-shaped isomer, which is the second lowest-energy (SiC)\textsubscript{12} isomer according to our searches and has a relative energy of 0.54 eV. We find an ionization energy of 7.77 eV and that its geometry is hardly affected by the ionization process (max 0.02 \AA{} / bond ), which also explains the low energy difference (0.10 eV) between vertical and adiabatic ionization. 
We investigated two other isomers that result in ionization energies of 7.75 eV and 7.77 eV above the neutral GM.

\section{Summary and conclusions}
The vertical and adiabatic single ionization energies of (SiC)\textsubscript{n}, n=1$-$12, span over a range of 6.59$-$10.64 eV and have an overall decreasing trend with cluster size n. In an astrophysical context, this energy range sets constraints for neutral (SiC)\textsubscript{n}, n=1$-$12, clusters to exist and therefore confines the conditions (temperature, radiation field) under which SiC nucleation can occur.
The adiabatically optimized cation geometries indicate that the ionization predominantly takes place in the silicon-segregated part of t/he clusters. 
For n=2,3,4 the lowest-energy cation does not correspond to the structural analogue of the neutral GM candidate structures, but to a different geometry. For these sizes, the structural analogue cations of the neutral GM show potential energies 0.83-1.66 eV and are hence metastable with respect to lowest-energy cation. 
Moreover, the optimization of the neutral GM structural analogues for sizes n=3,4,5 leads to transition states (TSs) that are subsequently re-optimized and real minima are found.\\
Applying the kinetic gas theory to the presented range of ionization energies (6.59$-$10.64 eV) results in equivalent temperatures of $\sim$76500$-$123500 K. These temperatures are significantly higher than the characteristic temperatures in envelopes and winds of evolved stars, even when including non-equilibrium effects like shocks or other non-thermal heating processes.
However, considering photo-ionization instead of ionization by a high temperature gas,the situation is different. For planetary nebulae with effective temperatures of $\sim$ 30000 K, the radiation field peaks for photon energies in the range 6 to 10 eV \citep{Cami1180}. This photon energy range is almost superimposable with the range of (SiC)\textsubscript{n} ionization energies derived in this study. 
Therefore, we conclude that the ionization of (SiC)\textsubscript{n} clusters in low-mass AGB stars is viable, in particular towards the end of their lives,
and that photo-ionization is most likely its responsible process.

\begin{table}
\caption{Vertical and adiabatic ionization energies, E\textsubscript{n}\textsuperscript{vert} and E\textsubscript{n}\textsuperscript{adia}(in eV), and their difference $\Delta$E=E\textsubscript{n}\textsuperscript{vert}-E\textsubscript{n}\textsuperscript{adia}, of (SiC)\textsubscript{n} clusters versus cluster size n\label{table1}.\textit{LM}
corresponds to a Local Minimum cation geometry, \textit{GM} to a Gloabl Minimum cation geometry, and \textit{TS} to a Transition State, respectively.}
\centering
\vspace*{0.3cm} 
\noindent\begin{tabular}{l r r r} 
n & E\textsubscript{n}\textsuperscript{vert} & E\textsubscript{n}\textsuperscript{adia} & $\Delta$E \\
\hline
\rule{0pt}{2ex}

1      & 10.64 & 10.64 & 0.00 \\
2 \textit{LM} & 10.01 &  9.87 & 0.14 \\
2 \textit{GM} &  8.27 &  8.21 & 0.07 \\
3 \textit{TS} &  9.34 &  8.95 & 0.39 \\
3 \textit{LM} &  9.34 &  8.77 & 0.57 \\
3 \textit{GM} &  8.17 &  7.94 & 0.22 \\ 
4 \textit{TS} &  8.78 &  8.63 & 0.15 \\
4 \textit{LM} &  8.78 &  8.23 & 0.55 \\
4 \textit{GM} &  7.62 &  7.36 & 0.27 \\   
5 \textit{TS} &  7.94 &  7.75 & 0.18 \\
5 \textit{GM} &  7.94 &  7.61 & 0.33 \\
6      &  7.90 &  7.46 & 0.44 \\
7      &  7.56 &  7.25 & 0.31 \\
8      &  7.22 &  6.85 & 0.37 \\
9      &  7.34 &  6.96 & 0.38 \\
10     &  7.20 &  6.95 & 0.25 \\
11     &  6.82 &  6.64 & 0.18 \\
12     &  6.69 &  6.59 & 0.10 \\                    
\end{tabular}
\end{table}






\section*{Conflict of Interest Statement}
The authors declare that the research was conducted in the absence of any commercial or financial relationships that could be construed as a potential conflict of interest.



\section*{Funding}
We aknowledge support from the ERC Consolidator Grant "AEROSOL - Astrochemistry of Old Stars: direct probing of unique chemical laboratories 2016-2020 (PI Leen Decin)", Project number 646758.

\section*{Acknowledgments}
We acknowledge the CINECA award under the ISCRA initiative, for the availability of high performance computing resources and support.



\bibliographystyle{frontiersinSCNS_ENG_HUMS} 
\bibliography{clustersic}




\end{document}